\documentclass{JHEP3}

\usepackage{graphicx}
\usepackage{verbatim}
\usepackage{epsf}
\usepackage{latexsym}
\usepackage{amsmath,amsfonts,amssymb,amsthm}

\def\bseq{\begin{subequation}}  
\def\eseq{\end{subequation}}
\def\bsea{\begin{subeqnarray}}  
\def\esea{\end{subeqnarray}}

\def\Tilde#1{\widetilde{#1}}                    

\newcommand{\bbox}{\lower.2ex\hbox{$\Box$}}

\newcommand{\beq}{\begin{equation}}
\newcommand{\eeq}{\end{equation}}
\newcommand{\bea}{\begin{eqnarray}}
\newcommand{\eea}{\end{eqnarray}}
\newcommand{\ena}{\end{eqnarray}}

\newcommand {\non}{\nonumber}

\newcommand{\Tr}{{\rm Tr}}
\renewcommand{\(}{\left(}
\renewcommand{\)}{\right)}

\newcommand{\be}{\begin{equation}}
\newcommand{\ee}{\end{equation}}

\newcommand{\firr}[1]{{}^{{\rm Irr}}\!{\cal F}^{\flat}_{#1}}

\preprint{LPTENS-09/07}

\title{\begin{center} 
3D Seiberg-like Dualities and M2 Branes
\end{center}}
\author{Antonio Amariti$^{1,a}$, Davide Forcella$^{2,b}$, Luciano Girardello$^{1,c}$ 
and Alberto Mariotti$^{3,d}$ 
\\ ~
\\
$^1$Dipartimento di Fisica, Universit\`a di Milano Bicocca\\
and \\
INFN, Sezione di Milano-Bicocca,\\ 
piazza della Scienza 3, I-20126 Milano, Italy\\
\\
$^2$ 
Laboratoire de Physique Th\'eorique 
de l'\'Ecole Normale Sup\'erieure \\
and  \\
CNRS UMR     8549 \\ 
  $24$ Rue Lhomond Paris $75005$, France\\
\\
$^3$
Theoretische Natuurkunde, Vrije Universiteit Brussel \\
and\\
The International Solvay Institutes\\ 
Pleinlaan 2, B-1050 Brussels, Belgium\\
 ~~\\
  $^a$\email{antonio.amariti@mib.infn.it} ~~~
  $^b$\email{forcella@lpt.ens.fr} ~~~
$^c$\email{luciano.girardello@mib.infn.it} ~~~
$^d$\email{alberto.mariotti@vub.ac.be} \\
}

\abstract{
We investigate features of duality in three dimensional
$\mathcal{N}=2$ Chern-Simons matter theories
conjectured to describe M2 branes at toric Calabi Yau
four-fold singularities. For 3D theories with non-chiral
4D parents we propone a Seiberg-like duality
which turns out to be a toric duality. For theories with chiral
4D parents we discuss the conditions under which that Seiberg-like duality
leads to toric duality. We comment on such duality in 3D theories without
4D parents.
}

\begin{document}

\section{Introduction}
Different descriptions of the same physical phenomenon usually provide a
better understanding of the phenomenon itself. AdS/CFT correspondence
and Seiberg duality are two famous examples.  In the AdS$_5$/CFT$_4$
case it happens that to a single geometry correspond different UV field
theory descriptions.  This phenomenon was called Toric Duality in
\cite{Feng:2000mi}, analyzed in \cite{Feng:2001xr,Feng:2002zw} and
identified as a Seiberg duality in \cite{Beasley:2001zp,Feng:2001bn}.
Due to the difficulties to understand the field theory living on M$2$
branes the AdS$_4$/CFT$_3$ correspondence was less mastered.  An
important step in this direction has been the realization of the importance
of Chern-Simons interactions \cite{Schwarz:2004yj} and the subsequent
construction of $\mathcal{N}=8$ Chern-Simons matter field theories in
\cite{Bagger:2006sk,Gustavsson:2007vu,Bagger:2007jr,Bagger:2007vi,
Gustavsson:2007vu,Gustavsson:2008dy,rams,Mukhi:2008ux}.
A breakthrough toward the explicit realization of the AdS$_4$/CFT$_3$
correspondence was done in \cite{Aharony:2008ug} where the authors propose 
$U(N)_k
\times U(N)_{-k}$ CS matter theories, with $\mathcal{N}=6$
supersymmetry, to be the low energy theories of $N$ M2-branes at the
$\mathbb{C}^4/\mathbb{Z}_k$ singularities.  Afterwards, this
construction has been extended to many others CS matter theories with
a lower amount of supersymmetries
\cite{
Benna:2008zy,hl3p,Hosomichi:2008jb,Schnabl,
Martelli:2008si,Hanany:2008cd,Hanany:2008fj,
Jafferis:2008qz,Gaiotto:2009mv,Imamura:2008nn,Imamura:2008ji,
Hanany:2008gx,Franco:2008um}.

A large and interesting, but still very peculiar, class of
AdS$_4$/CFT$_3$ pairs is realized by M$2$ branes at Calabi Yau
four-fold toric singularities 
\cite{Martelli:2008si,Hanany:2008cd,Hanany:2008fj,Hanany:2008gx,Franco:2008um,Ueda:2008hx,Imamura:2008qs}.
The
low energy theories are proposed to be a special kind of
$\mathcal{N}=2$ Chern-Simons matter theories.  It was soon realized
\cite{Hanany:2008fj} that, as in the AdS$_5$/CFT$_4$ case, different
$\mathcal{N}=2$ Chern-Simons matter theories can be associated with the
same Calabi Yau fourfold geometry.  The phenomenon of toric duality
reappears in the AdS$_4$/CFT$_3$ correspondence, but in a much general
context. Indeed, contrary to the four dimensional case, in three
dimensions one find models with different numbers of gauge group
factors to describe the same IR physics (for example mirror symmetry
pairs \cite{Kapustin:1999ha}). In the
literature some very specific
pairs of dual field theories were constructed.  A step was done
in \cite{Aharony:2008gk,Giveon:2008zn} where  a sort of
Seiberg like duality for three dimensional Chern-Simons matter theory
was proposed\footnote{Seiberg duality for 3D gauge theories were previously
studied in \cite{Aharony:1997bx,Karch:1997ux}.
}.
In the context of M$2$ branes at singularities, we can divide the set
of dualities in the ones that change and in the ones that do not
change the number of gauge group factors. In this paper we will
call the second type of duality Seiberg-like toric duality.  

$\mathcal{N}=2$ Chern-Simons matter
theories for M$2$ branes at singularities are typically described by a
quiver with an assignment of Chern-Simons levels $k_i$ and a
superpotential in a way similar to the gauge theories for D$3$ branes
at Calabi Yau three-fold singularities. Indeed a class of
$\mathcal{N}=2$ three dimensional theories can be simply obtained from
four dimensional $\mathcal{N}=1$ quivers with superpotential in the
following way: rewrite the theory in three dimensions, change the
$SU(N)$ gauge factors to $U(N)$ factors, disregard the super
Yang-Mills actions and add a super Chern-Simons term for every factors.
We will say that these three dimensional theories have a four
dimensional parent. Viceversa we will call theories without four
dimensional parents the three dimensional theories that cannot be
obtained in the way just explained \cite{Franco:2008um}.  

In this paper we investigate Seiberg-like toric dualities for (2+1)
dimensional $\mathcal{N}=2$ Chern-Simons matter theories associated
with M$2$ branes at Calabi Yau four-fold toric singularities.  Using a
generalization of the forward algorithm for D3 branes
\cite{Hanany:2008gx} we analyze a particular branch of the moduli
space that is supposed to reproduce the transverse four-fold Calabi
Yau singularity.  We identify a set of Seiberg-like toric dualities
for three dimensional Chern-Simons quiver theories.  

For theories with
four dimensional parents one could try to simply extend the four
dimensional Seiberg duality to the three dimensional case.
In fact, three dimensional theories share the same Master Spaces
\cite{Forcella:2008bb,Forcella:2008eh} of their four dimensional
parents. In the map between four and three dimensions a direction of
the Master Space become a direction of the physical Calabi Yau
four-fold.
Unfortunately, it turns out that an arbitrary assignment of 2+1
dimensional Chern-Simons levels does not in general commute with 3+1
dimensional Seiberg duality \cite{Hanany:2008fj}.
In fact it was shown in \cite{Forcella:2008ng} that the Master
Space for four dimensional Seiberg dual theories are not in general
isomorphic. Actually it seems that three dimensional CS theories with
chiral four dimensional parents do not admit a simple
generalization of the three dimensional SQCD 
Seiberg duality as it
happens in the four dimensional case.

Here, we first analyze non chiral three dimensional CS theories with
(3+1)d parents. Using 
a type $IIB$ brane realization,
we propose a Seiberg like duality, with a precise prescription
for the transformation of the CS levels and the gauge groups factors. 
We then check that 
this proposed Seiberg like duality 
is indeed a toric duality, namely that the two dual theories are
associated with M$2$ branes probing the same Calabi Yau four-fold
singularity.

We try to simply extend to chiral CS theories with (3+1)d parents  
the rules that we have found for the non chiral theories.  
For chiral four dimensional theories the Master Space is
not isomorphic among Seiberg dual phases \cite{Forcella:2008ng}.  
This fact presumably puts
constraints on the duality transformations for the three dimensional
case. We find indeed difficulties for a straightforward realization of
Seiberg like toric dualities for 2+1 dimensional Chern-Simons matter
theories with four dimensional chiral parents.

However, by analyzing several examples, we find
a rule for the assignments of the CS levels
such that
toric duality still holds among Seiberg like dual phases.

We finally give some examples of Chern-Simons theories without four
dimensional parents. In this case there is no immediate insight from the four
dimensions, but we show that the duality proposed for the chiral
theories works also for theories without a 3+1 parents.

Our analysis is a first step to the study of Seiberg-like toric
dualities in the context of M$2$ branes. We tried to use the intuition
from the non-chiral case and to leave as arbitrary as possible the values
of the Chern-Simons levels. It is reasonable that more general
transformation rules exist.  Moreover it would be nice to investigate
more general families of toric dualities, like the ones changing the
number of gauge group factors and the large limit for Chern-Simons
levels. We leave these topics for future investigations.
 
As we were finishing this paper, we were informed of \cite{Franco,Ami}
which discuss related topics.

\section{M2 branes and $\mathcal{N}=2$ Chern Simons
  theories}\label{secdavide}

As discussed in the introduction, supersymmetric Chern Simons theories
coupled to matter fields are good candidates to describe the low energy
dynamics of M2 branes \cite{Schwarz:2004yj,Aharony:2008ug}. In this
paper we are interested in M2 branes at Calabi Yau four fold toric
conical singularities. The authors of
\cite{Hanany:2008cd} proposed that the field theories
living on these M2 branes are $(2+1)$ dimensional $\mathcal{N}=2$
Chern Simons theories with gauge group $\prod_{i=1}^GU_i(N)$ with
bifundamental and adjoint matter fields.  The Lagrangian in
$\mathcal{N}=2$ superspace notation is:
\begin{equation} 
\label{forceculo2}
\Tr \( -i\sum_{a} k_a \int_0^1 dt V_a \bar{D}^{\alpha}( e^{t V_a} 
D_{\alpha}e^{-t V_a}) - \int d^{4} \theta 
\sum_{X_{ab}}
X_{ab}^{\dagger}e^{-V_a} X_{ab}e^{V_b} + \int d^{2} \theta W(X_{ab}) + c.c. \) 
\end{equation}
where $V_a$ are the vector superfields and $X_{ab}$ are bifundamental
chiral superfields.  The superpotential $W(X_{ab})$ satisfies the
toricity conditions: every field appears just two times: one time with
plus sign and the other time with minus sign. Since these theories
are conjectured to be dual to $M$ theory on $AdS_4 \times SE_7$, where
$SE_7$ is a seven dimensional Sasaki Einstein manifold, the moduli
space of these theories must contain a branch isomorphic to the
four-fold Calabi Yau real cone over $SE_7$: $\mathcal{M}_4=C(SE_7)$.
To study the moduli space we need to find the vanishing conditions for
the scalar potential.  The scalar potential is:
\begin{equation} 
  \Tr \( -4\sum_{a} k_a \sigma_a D_a + \sum_{a} D_a \mu_a(X) 
-\sum_{X_{ab}}|\sigma_a X_{ab} - X_{ab} \sigma_b |^2 
-\sum_{X_{ab}}|\partial_{X_{ab}}W|^2 \) \nonumber
\end{equation}
where $\mu_a(X)= \sum_b X_{ab} X_{ab}^{\dagger}-\sum_c
X_{ca}^{\dagger} X_{ca}+ [X_{aa}, X_{aa}^{\dagger}]$, $\sigma_a$ and
$D_a$ are scalar components of the vector superfield $V_a$, and with
abuse of notation $X_{ab}$ is the lowest scalar component of the
chiral superfield $X_{ab}$.  The moduli space is the zero locus of
the scalar potential and it is given by the equations:
\begin{equation}\label{vac}
\partial_{X_{ab}}W=0, \qquad \sigma_a X_{ab} - X_{ab} \sigma_b = 0, 
\qquad \mu_a(X)= 4 k_a \sigma_a. 
\end{equation}
In \cite{Hanany:2008cd} it was shown that if 
\be
\label{sumk}
\sum_{a} k_a = 0 \ee then the moduli space contains a branch
isomorphic to a four-fold Calabi Yau singularity.  This branch is
interpreted as the space transverse to the M2 branes.  Let us start
with the abelian case in which the gauge group is $U(1)^{G}$. We are
interested in the branch in which all the bifundamental fields are
generically different from zero.  In this case the solution to the
first equation in (\ref{vac}) gives the irreducible component of the
master space $\firr{~}$ \cite{Forcella:2008bb,Forcella:2008eh}. The
second equation in (\ref{vac}) imposes
$\sigma_{a_1}=...=\sigma_{a_G}=\sigma$. The last equation in
(\ref{vac}) are $G$ equations; the sum of all the equations gives
zero and there are just $G-1$ linearly independent
equations. The remaining $G-1$ equations can be divided in one along
the direction of the Chern Simons levels, and $G-2$ perpendicular to the
direction of the
Chern Simons levels.  The first equation fixes the value of
the field $\sigma$ while the other $G-2$ equations looks like
$\mu_i(X)=0$ and can be imposed, together with their corresponding
$U(1)$ gauge transformations, modding $\firr{~}$ by the complexified
gauge group action $(\mathbb{C}^*)^{G-2}$. The equation fixing the
field $\sigma$ leaves a $\mathbb{Z}_k$ action with
$gcd(\{k_\alpha\})=k$ by which we need to quotient to obtain the
moduli space. In the following we will take 
$gcd(\{k_\alpha\})=1$.  Summarizing, the branch of the moduli space we
just analyzed is: \be \mathcal{M}_4=\firr{~} \diagup H \ee where $H$
is the $(\mathbb{C}^*)^{G-2}$ kernel of \be
\label{DtermC}
C=\left(
\begin{array}{cccccc}
1&1&1&1&1&1\\
k_1&k_2&\dots&\dots&k_{G-1}&k_G
\end{array}
\right) \ee $\firr{~}$ is a $G+2$ dimensional toric Calabi Yau cone
\cite{Forcella:2008bb,Forcella:2008eh} and the vectors of charges in
$H$ are traceless by construction; it implies that $\mathcal{M}_4$ is
a four dimensional Calabi Yau cone and it is understood as the
transverse space to the $M2$ branes.  Following the same procedure in
the non abelian case it is possible to see that the moduli space
contains the $N$-times symmetric product of $\mathcal{M}_4$, and it is
interpreted as the the
transverse space to a set of $N$ BPS M2 branes.   \\

It is quite generic that a specific Calabi Yau four-fold is a branch
of the moduli space of apparently completely different $\mathcal{N}=2$
Chern-Simons theories. This fact it is called toric duality. We want
to systematically study $\mathcal{M}_4$ for some set of Chern Simons
theories and see if it is possible to find examples of toric dual
pairs. To do this we will use the algorithm proposed in
\cite{Hanany:2008gx}.

\subsection{An Algorithm to compute $\mathcal{M}_4$}

Let us review the algorithm proposed in \cite{Hanany:2008gx} to
compute $\mathcal{M}_4$.  We consider an $\mathcal{N}=2$ Chern-Simons
theory described in the previous section with gauge group $U(1)^{G}$,
and with the following constraints on the Chern-Simons levels:
\be \label{vincoli}
\sum k_a = 0
\ \ \ \ \ \ \ 
gcd(\{k_a\})=1
\ee
To compute $\mathcal{M}_4$ we need three matrices: the incidence
matrix $d$, the perfect matching matrix $P$, and the Chern-Simons
levels matrix $C$. $d$ contains the charges of the chiral fields under
the gauge group $U(1)$ factors of the theory, and can be easily
obtained from the quiver. $P$ is a map between the gauge linear sigma
model variables and the chiral fields in the Chern-Simons theory.  It
can be obtained from the superpotential of the theory and we refer the
reader to \cite{Hanany:2008gx, Franco:2008um} 
for explanations.  Summarizing, the determination
of the field theory contains the three matrices $d$, $P$, $C$.  They
are defined respectively by the gauge group representations of the
chiral fields, the chiral fields interactions and the Chern Simons
levels.  Once we get these three matrices we can obtain the toric
diagram of $\mathcal{M}_4$.  From $P$ and $d$ we compute the matrix
$Q$.  It is the matrix of charges of the gauge linear sigma model
variables under the $U(1)^G$ gauge group: $d = Q \cdot P^T$.  
From
$Q$ and $C$ we construct the charge matrix $Q_D=\hbox{ker}(C) \cdot
Q$. 
We denote with $K \equiv \hbox{ker}(C)$.
From $P$ we get the charge matrix $Q_F$: $Q_F=\hbox{ker}(P^{T})$.
Once we have $Q_D$ and $Q_F$ we combine them in the total charge
matrix $Q_t$: 
\be Q_{t}= \left(
\begin{array}{c}
Q_D\\
Q_F
\end{array}
\right)
\ee
The toric diagram of $\mathcal{M}_4$ is given by the kernel of
$Q_{t}$: 
\be 
G_t=(\hbox{ker}^*(Q_t))^T 
\ee 
where the columns of $G_t$ are
the vectors defining the toric diagram of $\mathcal{M}_4$. 
Note that, as pointed out in \cite{Hanany:2008gx},
we have to find the integer kernel, that we denote $\hbox{ker}^{*}$,
and not the nullspace of the charge matrix. Each row
of $G_t$ is reduced to a basis over the integer for every choice of the
CS levels.
We will see
this algorithm at work in the following sections.

\section{Non-Chiral Theories}

\begin{figure}[h!!!]
\begin{center}
\includegraphics[scale=0.55]{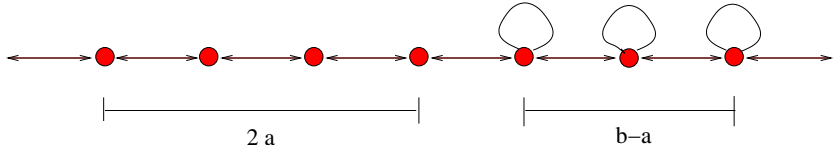}
\caption{The quiver for the generic $\Tilde{L^{aba}}_{\{k_i\}}$.}
\label{quivLaba}
\end{center}
\end{figure}

We consider non-chiral 3D $\mathcal{N}=2$ CS matter
quiver gauge theories which are
the three dimensional 
analog of 
the four dimensional $L^{aba}$ theories 
\cite{Benvenuti:2005ja,Butti:2005sw,Franco:2005sm}. 
and will be denoted as $\Tilde{L^{aba}}_{\{k_i\}}$.  
We say that a theory is non chiral 
if for every pair of gauge group factors $U(N)_i$ and $U(N)_{i+1}$ 
the number of bifundamental fields 
in the representation $(N_i,\bar{N}_{i+1})$ 
is the same as the number of bifundamental fields 
in the conjugate representation 
$(\bar{N_i},N_{i+1})$. 
Otherwise the theory is chiral. 
The quiver for the  $\Tilde{L^{aba}}_{\{k_i\}}$ is in figure \ref{quivLaba}, with
gauge groups $\prod_i U(N)_{k_i}$.
We label the nodes from left to right.
The action is
\bea \label{action}
S&&= \sum_{i} S_{CS} (k_i , V_i) \\
&&
+
\int d^{4} \theta \Tr 
\sum_i
( 
e^{-V_i} Q_{i,i+1}^{\dagger} e^{V_{i+1}} Q_{i,i+1}+
e^{V_i} Q_{i+1,i} e^{-V_{i+1}} Q_{i+1,i}^{\dagger}  ) +
\sum_j
X_{j,j}^{\dagger} e^{-2 V_{j}} X_{j,j}   \non \\
&&
+
\int d^{2} \theta
\sum_l (-1)^l   \Tr Q_{l-1,l}Q_{l,l+1}Q_{l+1,l}Q_{l,l-1} +
\sum_j  \Tr  Q_{j-1,j} X_{j,j} Q_{j,j-1}-Q_{j+1,j} X_{j,j} Q_{j,j+1} \non
\eea
where 
\be 
i=1, \dots a+b, \qquad  j=2a+1 \dots a+b, \qquad l=1 \dots 2a,
\ee
and $S_{CS} (k_i , V_i)$ is the first term in (\ref{forceculo2}).

\subsubsection*{Brane construction}
3D gauge theories can be engineered in type IIB 
string theory
as $D3$ branes suspended
among five branes \cite{Hanany:1996ie}.
For 3d CS theories the setup includes
$(p,q)5$ branes \cite{Kitao:1998mf,Bergman:1999na}.
Here we construct 
 $\mathcal{N}=2$
three dimensional $\Tilde{L^{aba}}_{\{k_i\}}$ 
CS theories, 
in analogy with 
the 4D 
construction \cite{Uranga:1998vf}.

\begin{figure}[h!!!]
\begin{center}
\includegraphics[scale=0.55]{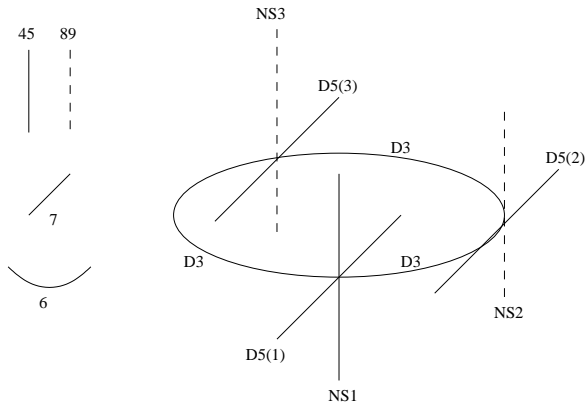}
\caption{Brane construction for $\Tilde{L^{121}}_{\{k_1,k_2,k_3\}}$. 
The D5 branes fill also the vertical
directions of the corresponding NS5.}
\label{L121}
\end{center}
\end{figure}

As an example we show in figure \ref{L121}
the realization of the $\Tilde{L^{121}}_{\{k_1,k_2,k_3\}}$ theory.
The generalization to the $\Tilde{L^{aba}}_{\{k_i\}}$ 
is straightforward.
We have the brane content resumed in Table \ref{tabellabrane}.
\begin{table}
\begin{center}
\begin{tabular}{c|c|c}
$\#$ &brane&directions\\
\hline
N& D3& 012 6 \\
\hline
1& NS$_1$& 012 3 45 \\
 \hline
1& NS$_2$& 012 3 89 \\
\hline
1&NS$_3$& 012 3 89 \\
\hline
$p_1$& D5$_1$&012 45 7 \\
 \hline
 $p_2$& D5$_2$& 012 7 89 \\
  \hline
 $p_3$& D5$_3$&  012 7 89 \\
\end{tabular}
\caption{Brane content for the $\Tilde{L^{121}}_{\{k_1,k_2,k_3\}}$ theory.}
\label{tabellabrane}
\end{center}
\end{table}
The NS branes and the corresponding D5 branes 
get deformed in ($1,p_i$) five branes at 
angles
$\tan \theta_i \simeq p_i$, obtaining
\begin{itemize}
\item N D3 brane along 012 6
\item ($1,p_1$) brane along 012 [3,7]$_{\theta_1}$ 45
\item ($1,p_2$) brane along 012 [3,7]$_{\theta_2}$ 89
\item ($1,p_3$) brane along 012 [3,7]$_{\theta_3}$ 89
\end{itemize}
Since the 
($1,p_2$) and the ($1,p_3$) branes
are parallel in the $89$ direction 
there is a massless adjoint 
field on the node $2$.
This brane system gives the three dimensional 
$\Tilde{L^{121}}_{\{k_1,k_2,k_3\}}$ CS theory.
 The Chern-Simons levels
are associated with the relative angle of the branes in the $[3,7]$
directions, i.e.
\be
\label{definizioK}
k_i=p_i-p_{i+1} \qquad i=1, \dots, a+b
\ee 
automatically satisfying (\ref{sumk}).
The gauge groups are all $U(N)$. 
Similar brane configurations have been studied in 
\cite{Jafferis:2008qz,Imamura:2008nn,Imamura:2008ji}
for $\mathcal{N}=3$
and/or non toric theories.

\subsection{Seiberg-like duality}
\label{SDfrombrane}
As in the four dimensional case \cite{Elitzur:1997fh}, 
we argue that
electric magnetic duality corresponds to
the exchange of two orthogonal ($1,p_i$) and ($1,p_{i+1}$) branes.
During this process,
$|p_i-p_{i+1}|=|k_{i}|$
D3 branes
are created \cite{Kitao:1998mf}.
Observe that, since the $p_i$ and $p_{i+1}$ 
of the dualized gauge group are exchanged,
this gives non trivial transformations also 
for the CS level of the neighbour nodes.
\begin{figure}[h!!!]
\begin{center}
\includegraphics[scale=0.4]{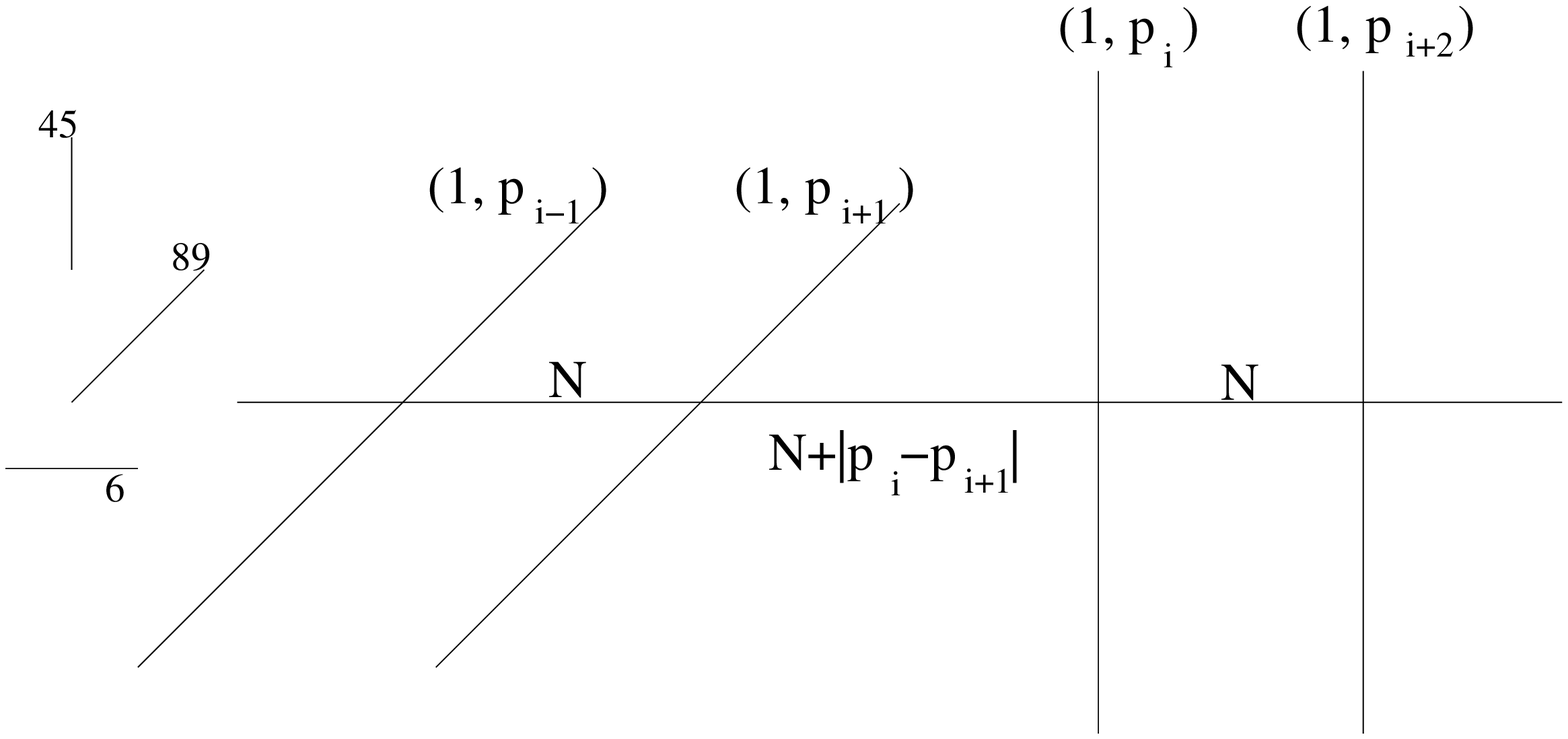}
\caption{Configuration after exchanging 
the position of the $(1,p_i)$ and $(1,p_{i+1})$ branes. 
The movement implies that $|p_{i}-p_{i+1}|$ $D3$ are created in the middle interval. 
The rank of the dualized group is $N+|p_{i}-p_{i+1}|=N+|k_i|$. The CS levels change
as $(p_{i-1}-p_i,p_i-p_{i+1},p_{i+1}-p_{i+2}) \to (p_{i-1}-p_{i+1},p_{i+1}-p_{i},p_{i}-p_{i+2}).$}
\label{scambio3}
\end{center}
\end{figure}

From the brane picture (see figure \ref{scambio3}) 
we obtain the 
rules for a Seiberg like duality on a node 
without adjoint fields
 in the $\Tilde{L^{aba}}_{\{k_i\}}$ quiver gauge theories.
 Duality on the $i$-th node gives
  \bea
  U(N)_{k_i} ~~~&\to& U(N+|k_i|)_{-k_i} \non \\
  \label{dualita}
 U(N)_{k_{i-1}} &\to& U(N)_{k_{i-1}+k_i}  \\
  U(N)_{k_{i+1}} &\to& U(N)_{k_{i+1}+k_i} \non 
 \eea
and
the field content and the superpotential 
changes
as in 4D Seiberg duality.

In the following we will verify that this is indeed a toric duality by computing and 
comparing the branch $\mathcal{M}_4$ of the moduli space (i.e. the toric diagram) 
of the two dual descriptions.

We observe that the Seiberg like duality 
 (\ref{dualita})
modifies the rank of the dualized gauge group,
introducing fractional branes.
This is a novelty of this $3d$ duality with respect to the $4d$ case
(see also 
\cite{Aharony:2008gk,Giveon:2008zn,Niarchos:2008jb}).

The $k$ fractional $D3$ branes are stuck between the 
five branes, so there is no moduli space associated with their motion. 
This is as discussed in \cite{Aharony:2008gk}, and
the same field theory argument can be repeated here.
The moduli space of the magnetic description is then
the $N$ symmetric product of the abelian moduli space.

The fractional branes can
break supersymmetry as a consequence of the s-rule 
\cite{Hanany:1996ie}.
Indeed it was suggested
in \cite{Witten:1999ds,Ohta:1999iv,Bergman:1999na} 
that for $U(l)_k$ YM-CS theories
supersymmetry is broken if $l>|k|$.
We notice from (\ref{dualita})
that in the moduli space of the magnetic description,
there is a pure $U(k)_{-k}$ YM-CS theory.
Thus the bound is satisfied and
supersymmetry is unbroken. 
However,
if we perform multiple dualities we can realize
configurations with several fractional branes on different nodes. 
At every duality we have to
control 
via s-rule
that supersymmetry is not broken. 
We leave a more thorough study of
these issues related to 
fractional branes for future investigation.

Finally, the duality proposed maps a theory with a weak coupling 
limit to a strongly coupled theory. 
Indeed if we define the $i$-th 
't Hooft coupling as $\lambda_i=N/k_i$, 
the original theory is weakly coupled for $k_i>> N$. 
In this limit the dual theory is strongly coupled 
since the i-th dual
't Hooft coupling is $\tilde \lambda_i= 1+ O(N/k_i)$.

\subsection{$\Tilde{L^{121}}_{\{k_1,k_2,k_3\}}$}
\begin{figure}[h!!!]
\begin{center}
\includegraphics[scale=0.5]{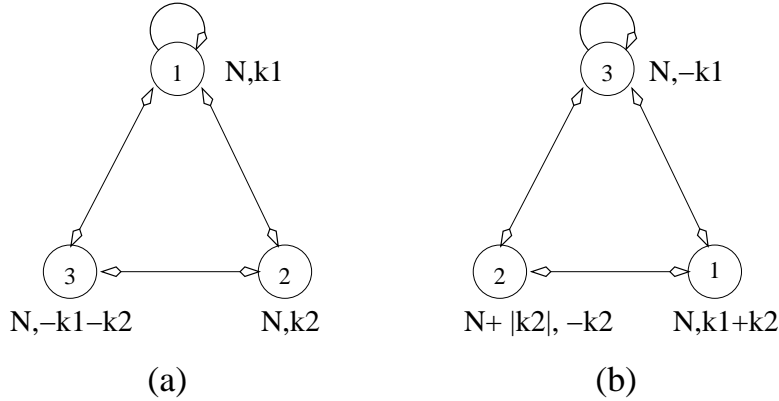}
\caption{Quiver, ranks and CS levels for the 
$\Tilde{L^{121}}_{\{k_1,k_2,k_3\}}$ in the two phases 
related by Seiberg like duality on node $2$. }
\label{SPPfig}
\end{center}
\end{figure}

The $\Tilde{L^{121}}_{\{k_1,k_2,k_3\}}$ is the first example 
that we study. 
The quiver is given in Figure \ref{SPPfig}.
The toric diagram that encodes the information about the classical
mesonic moduli space is computed with the techniques explained in
section \ref{secdavide}.  We extract the incidence matrix
$d_{G,i}$, where $G=1,\dots,N_g$ runs over the labels of the gauge
groups, and $i$ runs over the fields ($i=1,\dots,7$ for the $\Tilde{L^{121}}_{\{k_1,k_2,k_3\}}$, six
bifundamental and one adjoint)
\be
d=\left(
\begin{array}{ccccccc}
1&-1 &0 &0 &-1 &1 &0 \\
-1&1&1&-1&0&0&0\\
0&0&-1&1&1&-1&0
\end{array}
\right)
\ee
The matrix of the perfect matchings $P_{\alpha,i}$ is computed from
the determinant of the Kastelein matrix
\be \label{kasmat}
Kas = 
\left(\begin{array}{cc}
Q_{23}+Q_{32}& Q_{12}+Q_{21}\\
Q_{13}+Q_{31}&X_{11}
\end{array} \right)
\ee
If we order the fields in the determinant of
({\ref{kasmat}) we can build the matrix $P_{i,\alpha}$
where $\alpha=1,\dots,c$ is the number of perfect matchings,
that corresponds to the number of monomials of the $\det{Kas}$.
In this matrix we have $1$ if the $i$-th field
appears in the $\alpha$-th element of the determinant,
$0$ otherwise
\be
P = \left(
\begin{array}{cccccc}
1 & 0 & 0 & 0 & 1 & 0 \\
0 & 1 & 0 & 0 & 0 & 1 \\
0 & 0 & 0 & 1 & 0 & 0 \\
0 & 0 & 1 & 0 & 0 & 0 \\
0 & 1 & 0 & 0 & 1 & 0 \\
1 & 0 & 0 & 0 & 0 & 1 \\
0 & 0 & 1 & 1 & 0 & 0 
\end{array}
\right)
\ee
The matrix $Q$ that represents the charge matrix 
for the GLSM fields is obtained from the relation
$d_{G,i}=Q_{G,\alpha}\cdot P_{i,\alpha}^T$
\be
Q = \left(
\begin{array}{cccccc}
 1  & -1 &  0 & 0 &  0 & 0\\
 0  &  1 & -1 & 1 & -1 & 0\\
-1  &  0 &  1 &-1 &  1 & 0
\end{array}
\right)
\ee
The contribution of the $D$-terms to
the moduli space is given by quotienting 
by the $G-2$ FI parameters induced by the CS couplings.
These FI parameters are in the integer kernel of the matrix
of the CS level
\be
K=Ker\left(
\begin{array}{ccc}
1&1&1\\
k_1&k_2&-k_1-k_2
\end{array}
\right)
\ee
The $F$-term equations are encoded in the
matrix $Q_F=Ker(P)$. 
The Toric diagram is the kernel of the matrix obtained 
by combining $Q_{D}=K \cdot Q$
and $Q_F$. 
Acting with an $SL(4,Z)$ transformation
the toric diagram reads
\be \label{toricSPPdiag}
G_t=
Ker^{*}\left[
K
\cdot
Q,
Ker[P^T]
\right]=
\left(
\begin{array}{cccccc}
1&1&1&1&1&1\\
k_2&k_1+k_2&0&0&k_1+2k_2&0\\
1&1&0&1&2&0\\
0&0&1&1&0&0\\
\end{array}
\right)
\ee
This
system of vectors is co-spatial.
This is a CY condition that guarantees that the toric diagram 
lives on a three dimensional hypersurface in $Z_4$.

The last three rows of (\ref{toricSPPdiag})
defines the
 toric diagram for the three
dimensional Chern-Simons $\Tilde{L^{121}}_{\{k_1,k_2,k_3\}}$ 
toric quiver gauge theory. Note that
the toric diagram of the (3+1)d parent is recovered by
setting to zero the row with the CS levels.

We perform the Seiberg-like duality (\ref{dualita}) on node 2.  
The resulting theory is shown in figure \ref{SPPfig}(b).
The
$L^{121}$ four dimensional parent theory has only one toric phase.
The dual theory is in the same phase, thanks to the mapping among the
nodes ($1 \to 3, 2 \to 1, 3 \to 2$), see Figure \ref{SPPfig}. 
In the $\Tilde{L^{121}}_{\{k_1,k_2,k_3\}}$ we should also properly
map the CS levels in the two dual descriptions.
The transformation rules (\ref{dualita}) change the CS level
as in figure \ref{SPPfig}. Then we apply the same mapping
we used for the gauge groups.
After these steps the resulting $K$ matrix is
\be 
K_{dual}= 
Ker\left(
\begin{array}{ccc}
1&1&1\\
\tilde k_3& \tilde k_1&\tilde k_2
\end{array}
\right)
=
Ker\left(
\begin{array}{ccc}
1&1&1\\
k_3+k_2&k_1+k_2&-k_2
\end{array}
\right)
=
Ker\left(
\begin{array}{ccc}
1&1&1\\
- k_1&k_1+k_2&-k_2
\end{array}
\right)
\ee
where with $\tilde k_i$ we denote the CS level of the $i$-th node
in the dual phase.
Concerning the field content and the superpotential,
the dual theory is in the same phase 
than the starting theory.
Thus we use the same matrices $P,d,Q$ for the
computation of the moduli space.
The toric diagram is then computed with the usual algorithm.
Up to an $SL(4,Z)$ transformation it coincides with the same as
the one computed in the original theory (\ref{toricSPPdiag}).

In this example we have shown that the Seiberg like duality (\ref{dualita})
is a toric duality. Observe that the non trivial transformation
on the CS levels of (\ref{dualita}) are necessary for
the equivalence of the moduli spaces of the two phases.

\subsection{$\Tilde{L^{222}}_{\{k_i\}}$ }
The second example is the $\Tilde{L^{222}}_{\{k_i\}}$  theory.
The main difference is that this theory has 
two phases with a different matter content and
superpotential (see Figure \ref{quivL222}), 
obtained by dualizing node $2$. 
These phases are dual for the (3+1)d parents theory.
Here we show that the same holds in three dimensions
with the Seiberg like duality (\ref{dualita}).
\begin{figure}[h!!!]
\begin{center}
\includegraphics[scale=0.55]{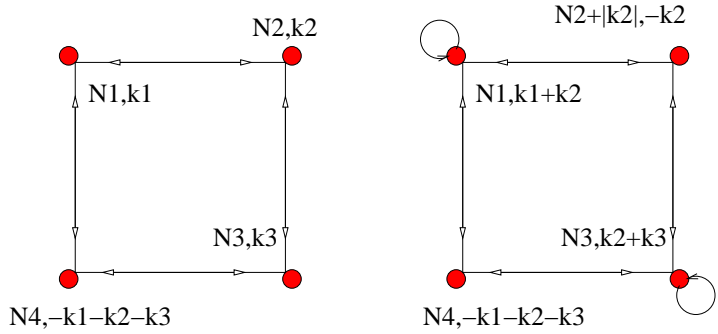}
\caption{The quivers for the two phases of $L^{222}$.}
\label{quivL222}
\end{center}
\end{figure}
The $P,d,Q$ and $K$ matrices
for the first phase are 
\be
d=\left(
\begin{array}{cccccccc}
1&-1 &0 &0 &0 &0 &-1 &1\\
-1&1&1&-1&0&0&0&0\\
0&0&-1&1&1&-1&0&0\\
0&0&0&0&-1&1&1&-1\\
\end{array}
\right)
%
\quad
P = \left(
\begin{array}{cccccccc}
0 & 0 & 0 & 0 & 1 & 0 & 1 & 0\\
0 & 0 & 0 & 0 & 0 & 1 & 0 & 1\\
1 & 0 & 1 & 0 & 0 & 0 & 0 & 0\\
0 & 1 & 0 & 1 & 0 & 0 & 0 & 0\\
0 & 0 & 0 & 0 & 1 & 1 & 0 & 0\\
0 & 0 & 0 & 0 & 0 & 0 & 1 & 1\\
1 & 1 & 0 & 0 & 0 & 0 & 0 & 0\\
0 & 0 & 1 & 1 & 0 & 0 & 0 & 0\\
\end{array}
\right)
\ee
\be
Q = \left(
\begin{array}{cccccccc}
 -1 &  0 &  1 & 0 &  1 & -1 & 0 &  0 \\
  1 & -1 &  0 & 0 & -1 &  1 & 0 &  0 \\
 -1 &  1 &  0 & 0 &  0 &  1 & 0 & -1 \\
  1 &  0 & -1 & 0 &  0 & -1 & 0 &  1 \\
\end{array}
\right)
%
\quad
K=Ker\left(
\begin{array}{cccc}
1&1&1&1\\
k_1&k_2&k_3&-k_1-k_2-k_3
\end{array}
\right)
\ee
The resulting toric diagram is
\be
\label{tdL222}
G_t=
\left(
\begin{array}{cccccccc}
1 & 1 & 1 & 1 & 1 & 1 & 1 & 1 \\
2 & 1 & 1 & 0 & 2 & 1 & 1 & 0 \\
1 & 1 & 1 & 1 & 0 & 0 & 0 & 0 \\
-k1-k_2&-k_1&k_2&0&k_2+k_3&k_2+k_3&0&0
\end{array}
\right)
\ee
By duality on node $2$ we obtain the inequivalent phase of $\Tilde{L^{222}}_{\{k_i\}}$ ,
see figure \ref{quivL222}.
The toric diagram is computed with new 
$d$, $P$, $Q$ and $K$ matrices
\be
d=\left(
\begin{array}{cccccccccc}
0 &  1 & -1 &  0 &  0 & 0 & 0 & 0 &-1 &1\\
0 & -1 &  1 &  1 & -1 & 0 & 0 & 0 & 0 &0\\
0 &  0 &  0 & -1 &  1 & 0 & 1 &-1 & 0 &0\\
0 &  0 &  0 &  0 &  0 & 0 &-1 & 1 & 1 &-1
\end{array}
\right)
%
\quad
P = \left(
\begin{array}{cccccccc}
0 & 0 & 0 & 0 & 1 & 1 & 1 & 1\\
1 & 0 & 1 & 0 & 0 & 0 & 0 & 0\\
0 & 1 & 0 & 1 & 0 & 0 & 0 & 0\\
0 & 0 & 0 & 0 & 1 & 0 & 1 & 0\\
0 & 0 & 0 & 0 & 0 & 1 & 0 & 1\\
1 & 1 & 1 & 1 & 0 & 0 & 0 & 0\\
0 & 0 & 0 & 0 & 1 & 1 & 0 & 0\\
0 & 0 & 0 & 0 & 0 & 0 & 1 & 1\\
1 & 1 & 0 & 0 & 0 & 0 & 0 & 0\\
0 & 0 & 1 & 1 & 0 & 0 & 0 & 0\\
\end{array}
\right)
\ee
\be
Q = 
\left(
\begin{array}{cccccccc}
0&-1&1&0&0&0&0&0\\
0&0&-1&1&0&0&1&-1\\
0&0&0&0&0&1&-1&0\\
0&1&0&-1&0&-1&0&1
\end{array}
\right)
%
\quad
K=Ker\left(
\begin{array}{cccc}
1&1&1&1\\
k_1+k_2&-k_2&k_3+k_2&-k_1-k_2-k_3
\end{array}
\right)
\ee
It results
\be
G_t=
\left(
\begin{array}{cccccccc}
1 & 1 & 1 & 1 & 1 & 1 & 1 & 1 \\
2 & 1 & 1 & 0 & 2 & 1 & 1 & 0 \\
-k1-k_2&-k_1&k_2&0&k_2+k_3&k_2+k_3&0&0\\
1 & 1 & 1 & 1 & 0 & 0 & 0 & 0 \\
\end{array}
\right)
\ee
which is equivalent to (\ref{tdL222}).

\subsection{The general $\Tilde{L^{aba}}_{\{k_i\}}$}

In the previous section we have seen two simple examples.
In this section we consider the generic case of non-chiral $\mathcal{N}=2$
toric three dimensional CS quiver gauge theories $\Tilde{L^{aba}}_{\{k_i\}}$ singularities.
Four dimensional theories based on these singularities
share the same toric diagram among the different Seiberg dual
phases. 
Here we show that two theories that are related by Seiberg like 
duality 
in three dimension (\ref{dualita}) share the same toric diagram.
Our argument is based on the algorithm
\cite{Ueda:2008hx} that extracts toric data of the CY four-fold by 
using brane tiling.

\subsubsection*{Toric diagrams from bipartite graphs}

Let us remind the reader that to every quiver describing a
four dimensional conformal field theory on D3 branes at Calabi Yau three-fold 
singularities it is possible to associate a bipartite diagram drawn on a 
torus. It is called tiling or dimer \cite{Hanany:2005ve,Franco:2005rj}, and it encodes all
the informations in the quiver and in the superpotential. To every face in the
dimer we can associate a gauge group factor, to every edge a bifundamental field and to 
every node a term in the superpotential. A similar tiling can be associated with 
three dimensional Chern-Simons matter theories living on M$2$ branes probing a 
Calabi Yau four-fold singularity simply adding a flux of Chern-Simons charge. 
The CS levels are described as a conserved flow
 on the quiver, or equivalently on the dimer.
To every edge we associate a flux of Chern-Simons charge
and the CS level of the gauge group is the sum of these
contributions taken with sign depending on the orientation of the arrow.

In this section we review the proposal of \cite{Ueda:2008hx} for the
computation of the moduli space of three dimensional 
CS toric quiver gauge theories. This
method furnishes the toric data from the bipartite graphs associated
with the quiver model and with the CS levels.

One has to choose a set of paths ($p_1,\dots,p_4$)on the dimer. 
The paths $p_1$ and $p_2$ correspond to the
$\alpha$ and $\beta$ cycles of the torus described by the dimer. The
path $p_4$ is a paths encircling one of the vertices.
One can also associate mesonic operators to these paths. These
operators correspond to the product of the corresponding
bifundamentals along the paths. For example the operator associated
on the $p_4$ path is a term of the superpotential.
The moduli space of the three dimensional theory requires also the
definition of the path $p_3$. This is a product of paths
corresponding to a closed flow of CS charges along the quiver.
We choose $p_3$ in the tiling of the $L^{aba}$ singularity
by taking a minimal closed path connecting the
bifundamentals from the first node to the $a+b$-th node of the
quiver. Then one associates the CS charge
\be \sum_{i=1}^{n} k_i\ee
to each bifundamental in this closed loop, with $n=1,\dots,a+b$. The
last charge is zero since it corresponds to sum of all the CS levels
in the theory. This conserved CS charges flow is represented on the
dimer by a set of oriented arrows connecting the faces of the dimer,
the gauge groups.

In \cite{Ueda:2008hx} it has been shown that the toric polytope of
$CY_4$ is given by the convex hull of all lattice point
$v^{\alpha}=(v_1^\alpha,\dots,v_4^\alpha)$, with
\be
\label{toricpolytope} v_i^\alpha=\langle p_i,D_\alpha \rangle
\ee
where $D_\alpha$ are the perfect matchings. The operation $\langle
.,.\rangle$ in (\ref{toricpolytope}) is the signed intersection
number of the perfect matching $D_\alpha$ with the path $p_i$. Note
that $v_4$ is always $1$, since there is always only one
perfect matching connected with the node encircled by $p_4$. 

\subsubsection*{Seiberg-like duality on $\Tilde{L^{aba}}_{\{k_i\}}$ and toric duality}
\begin{figure}[h!!!]
\begin{center}
\includegraphics[width=10cm]{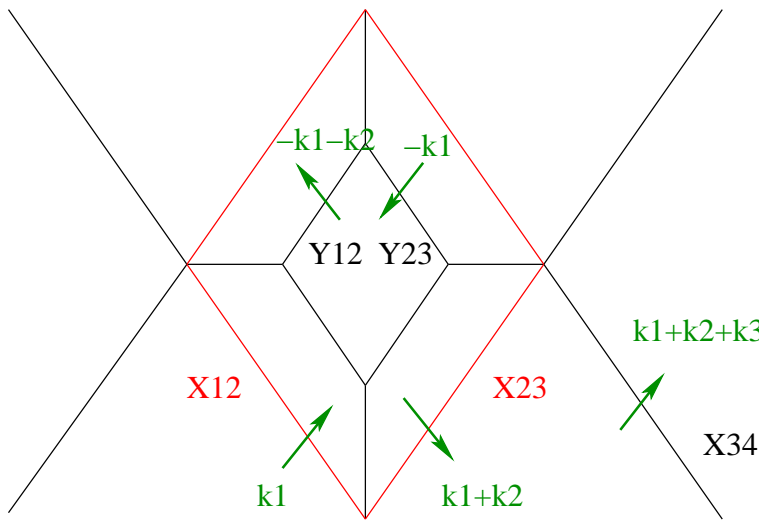}
\caption{Action of Seiberg duality on the dimer and modification of the 
CS flow.}
\label{Sduality}
\end{center}
\end{figure}
In subsection (\ref{SDfrombrane}) we argued that 
the action of Seiberg duality on the field content 
and on the superpotential is the same as in four dimensions. 
The only difference is the change of the CS levels 
associated with the groups involved in the duality (\ref{dualita}).

In an $L^{aba}$ theory, if duality is performed on node
$N_2$, the levels become
\bea
\label{newlevels}
&&k_1 \rightarrow k_1+k_2 \nonumber \\
&&k_2 \rightarrow -k_2 \nonumber \\
&&k_3 \rightarrow k_3+k_2
\eea
The action of Seiberg duality (\ref{dualita}) 
not only modifies the dimer as in
the 4d, but also the CS levels of the gauge groups.
This changes the path $p_3$ in the dual description. Many $SL(4,Z)$
equivalent choices are possible. Among them we select the $p_3$
path as in figure \ref{Sduality}.
We associate
a charge $l_1$ to $Y_{12}$
and $l_2$ to $Y_{23}$, but with the opposite arrows that before. 
The
values of the charges $l_1$ and $l_2$ are derived from
(\ref{newlevels}) and are 
\be l_1 = -k_1-k_2 \ \ \ \ \ l_2=-k_1 \ee
The field $X_{34}$ is not involved in this duality and it contributes
to the CS flow with the same charge $k_1+k_2+k_3$ in both phases.

We claim that the two theories share the same toric diagram.
For the proof of this relation 
it is useful to distinguish
two sector of fields from which all the perfect matching are built.
In Figure \ref{PMfigure} we separated these two sectors for the
electric and magnetic phase of an $L^{aaa}$ theory (the same
distinction is possible in a generic $L^{aba}$ theory). Every perfect
matching is built by choosing in these sets only one field
associated with each vertex.
For example in Figure \ref{PMfigure}(a) every perfect matching 
is a set of blue lines chosen such that every vertex is involved
only once.

\begin{figure}
\begin{center}
\begin{tabular}{cc}
\includegraphics[width=6cm]{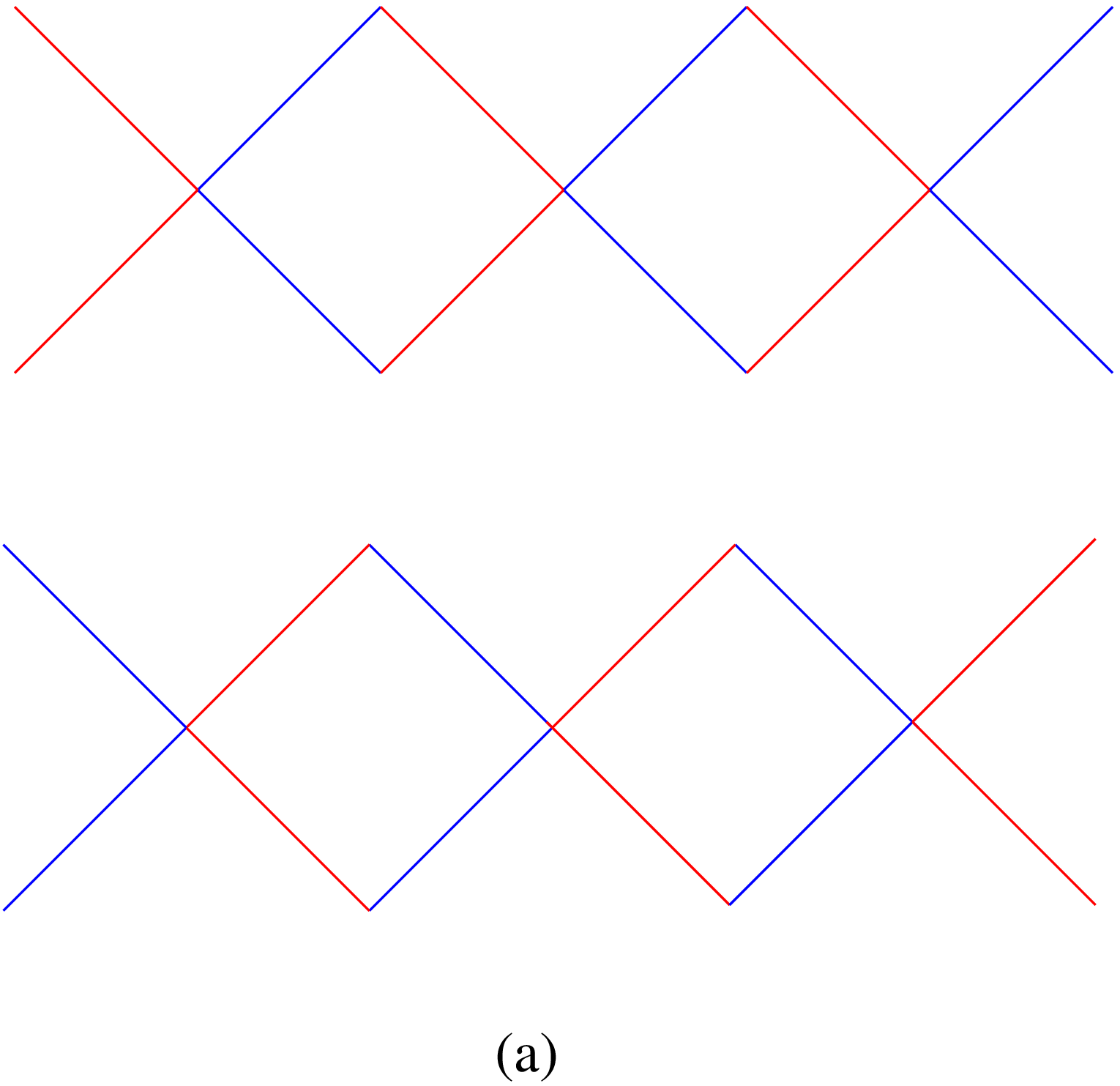}
&
\includegraphics[width=8.5cm]{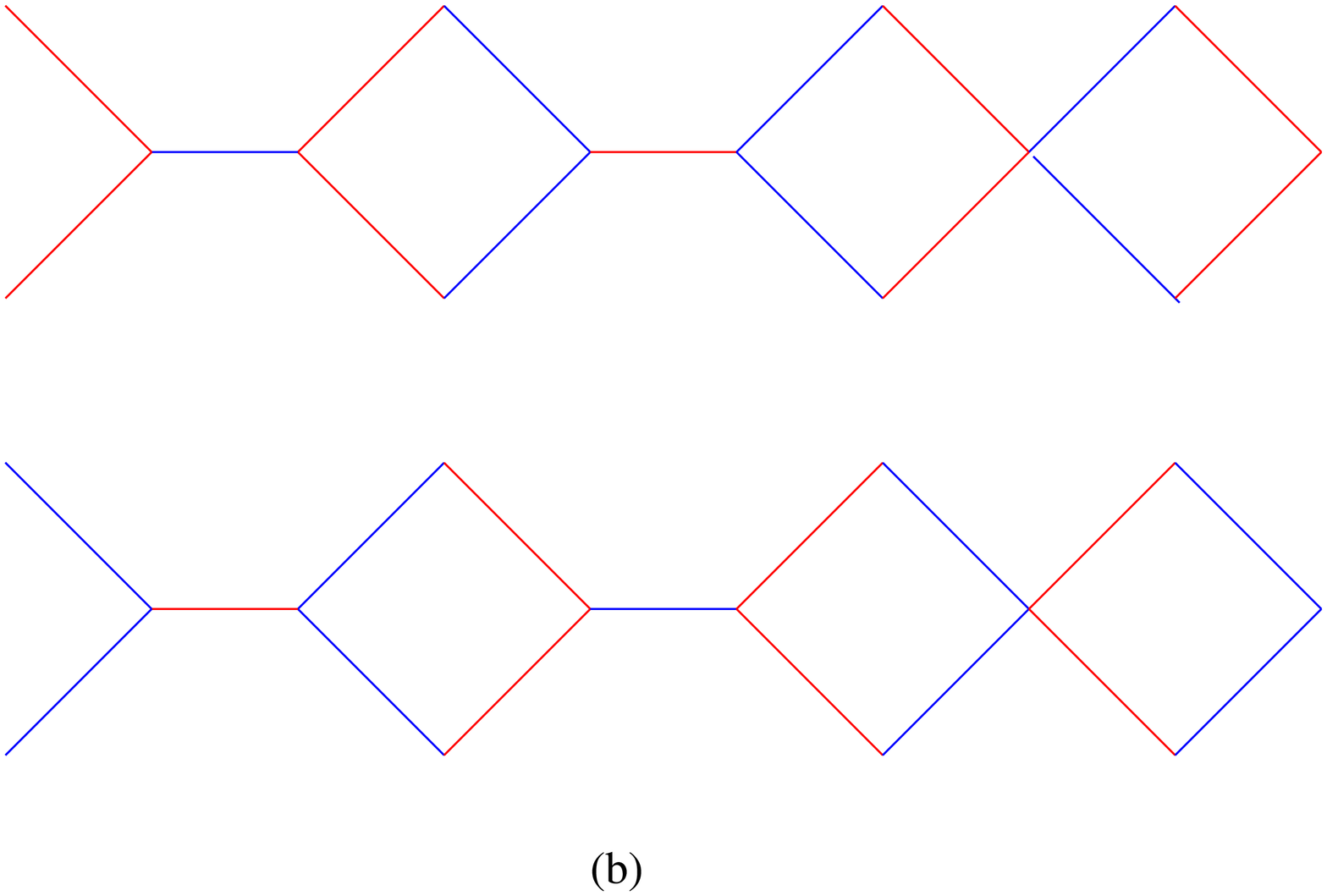}\\
\end{tabular}
\caption{Different sectors of fields that generate all perfect matching in the 
(a) electric and (b) magnetic theory}
\label{PMfigure}
\end{center}
\end{figure}

The paths $p_1$ and $p_2$ are shown in figure \ref{p_ichoice}
for the electric and the magnetic phase.
\begin{figure}
\begin{center}
\begin{tabular}{cc}
\includegraphics[width=7.3cm]{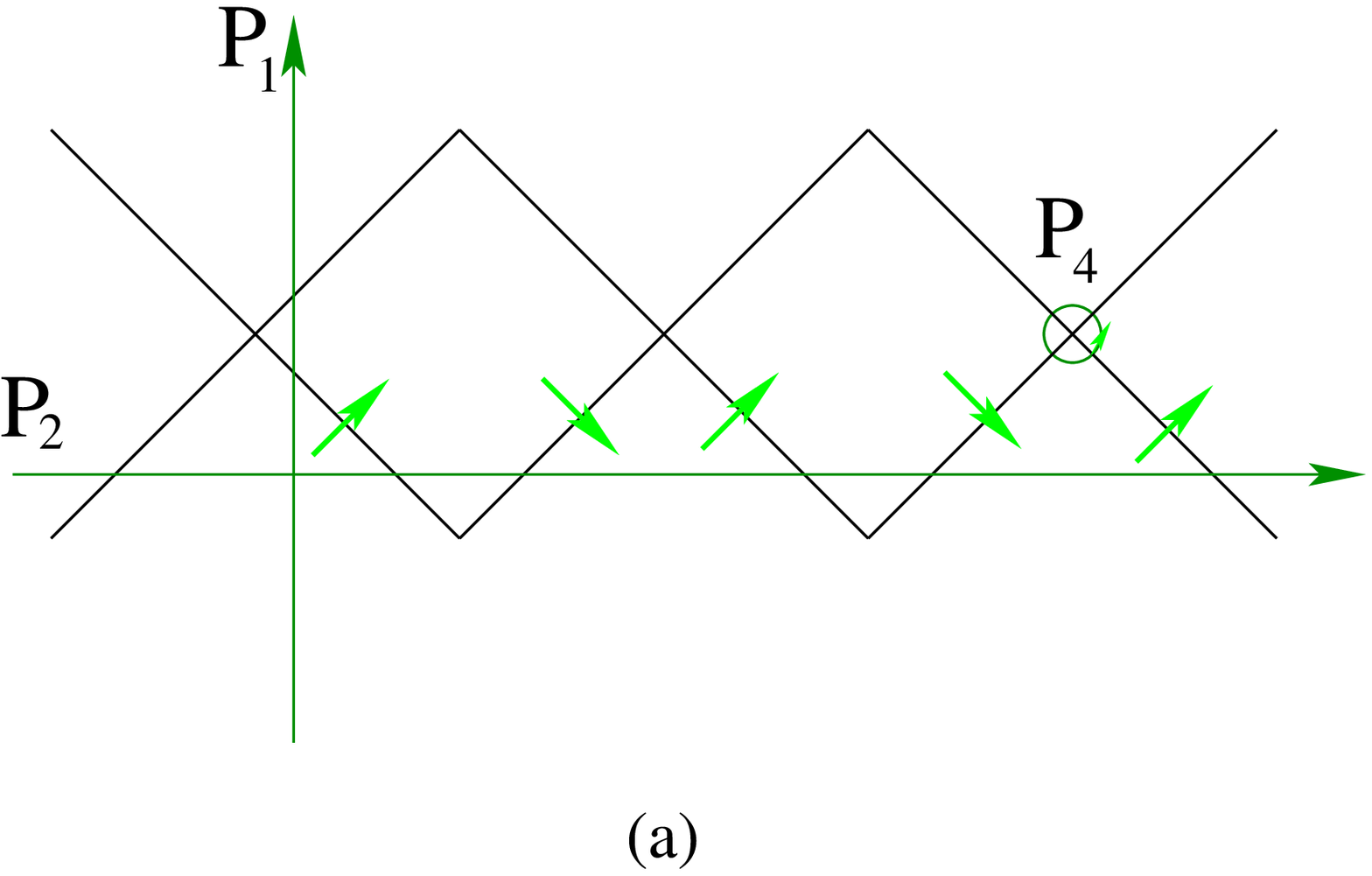}
&
\includegraphics[width=8.7cm]{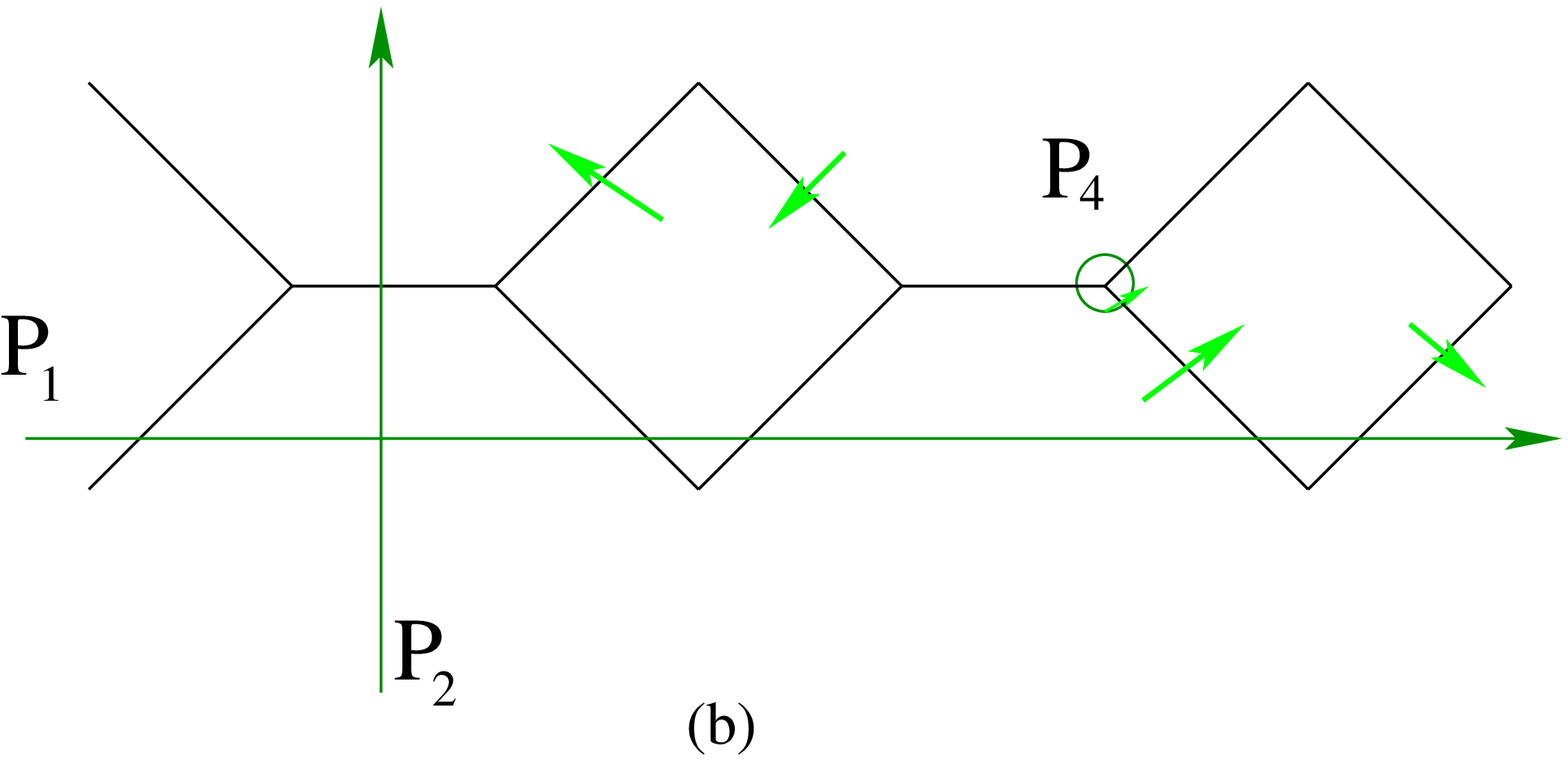}\\
\end{tabular}
\caption{Paths $p_i$ in the two dual versions of the theory}
\label{p_ichoice}
\end{center}
\end{figure}
The intersection numbers $\langle p_1,D^{\alpha} \rangle$
and $\langle p_2,D^{\alpha}\rangle$ give the same bidimensional 
toric diagram of the associated four dimensional $L^{aba}$
theories, up to 
an overall
$SL(3,Z)$ translation.
This is shown by mapping the perfect matching 
in the two description. This mapping is done with a prescription 
on the choice of the fields in the 
perfect matching of the dual description.
If duality is performed on node $N_i$,
 the field $X_{i-1,i}$, $X_{i,i-1}$,
$X_{i,i+1}$ and $X_{i+1,i}$ are respectively mapped in the fields
$Y_{i,i+1}$, $Y_{i+1,i}$, $Y_{i-1,i}$ and $Y_{i.i-1}$ of the dual theory.
This prescription gives a $1-1$ map of each point of the 2d toric diagrams 
of the two dual theories.

The whole diagram for the three dimensional $\Tilde{L^{aba}}$ theory
is obtained by considering the intersection numbers 
$\langle p_3,D^{\alpha} \rangle$, which give the third component of the
vectors $v^{\alpha}$.
The path $p_3$ corresponds to the flux of CS
charges from the group $N_1$ to the group $N_g$, where the last
arrow is omitted since it carries zero charge. In the dual theory
the path $p_3$ changes as explained above, and as
we show in Figure \ref{p_ichoice} for duality on a node labelled by 2. 
Note that only
the arrows connected with the dualized gauge group change.

With this choice of $p_3$ and using the basis of perfect matching
we prescribed, the intersection numbers $\langle p_3,D^{\alpha} \rangle$
coincide in the two phases for every point of the 2d toric diagram.
 This is shown by associating the relevant part of
the path $p_3$ to each sector of perfect matching 
as in Figure \ref{PM_and_p3}. The arrow that carries charge $k_1$ in the
electric theory corresponds to the arrow with charge $-k_1$ in the
magnetic theory. Its contribution to the moduli space remains the
same, since also the orientation of this arrow is the opposite. The
same happens for the arrow carrying charge $k_1+k_2$.

Thus the 3d toric diagrams of the two dual theories are the same. 
We conclude that the action of three dimensional 
Seiberg like duality (\ref{dualita}) in the $\Tilde{L^{aba}}_{\{k_i\}}$ theories implies toric duality.

\begin{figure}
\begin{center}
\begin{tabular}{cc}
\includegraphics[width=5.5cm]{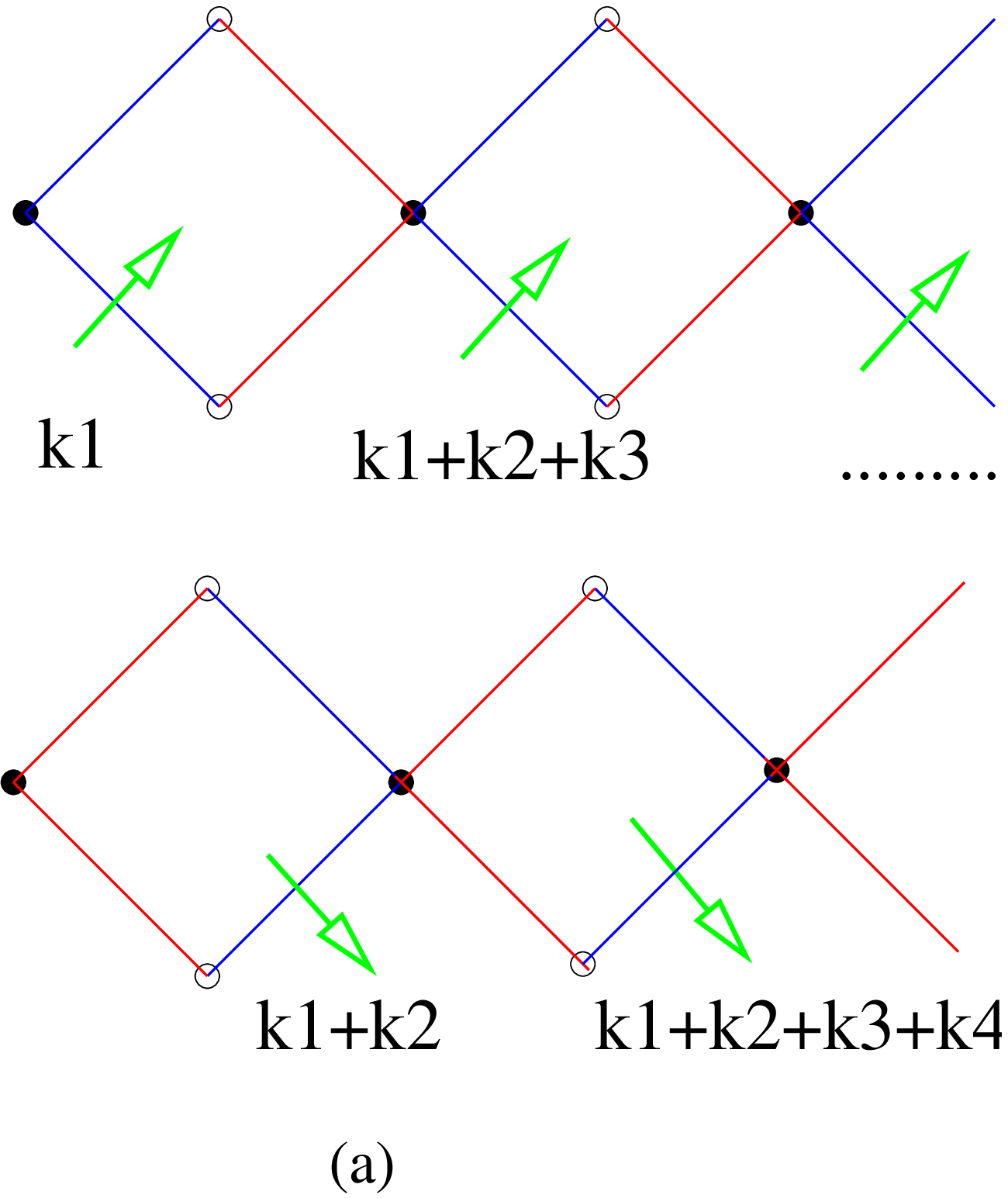}
&
\includegraphics[width=8.5cm]{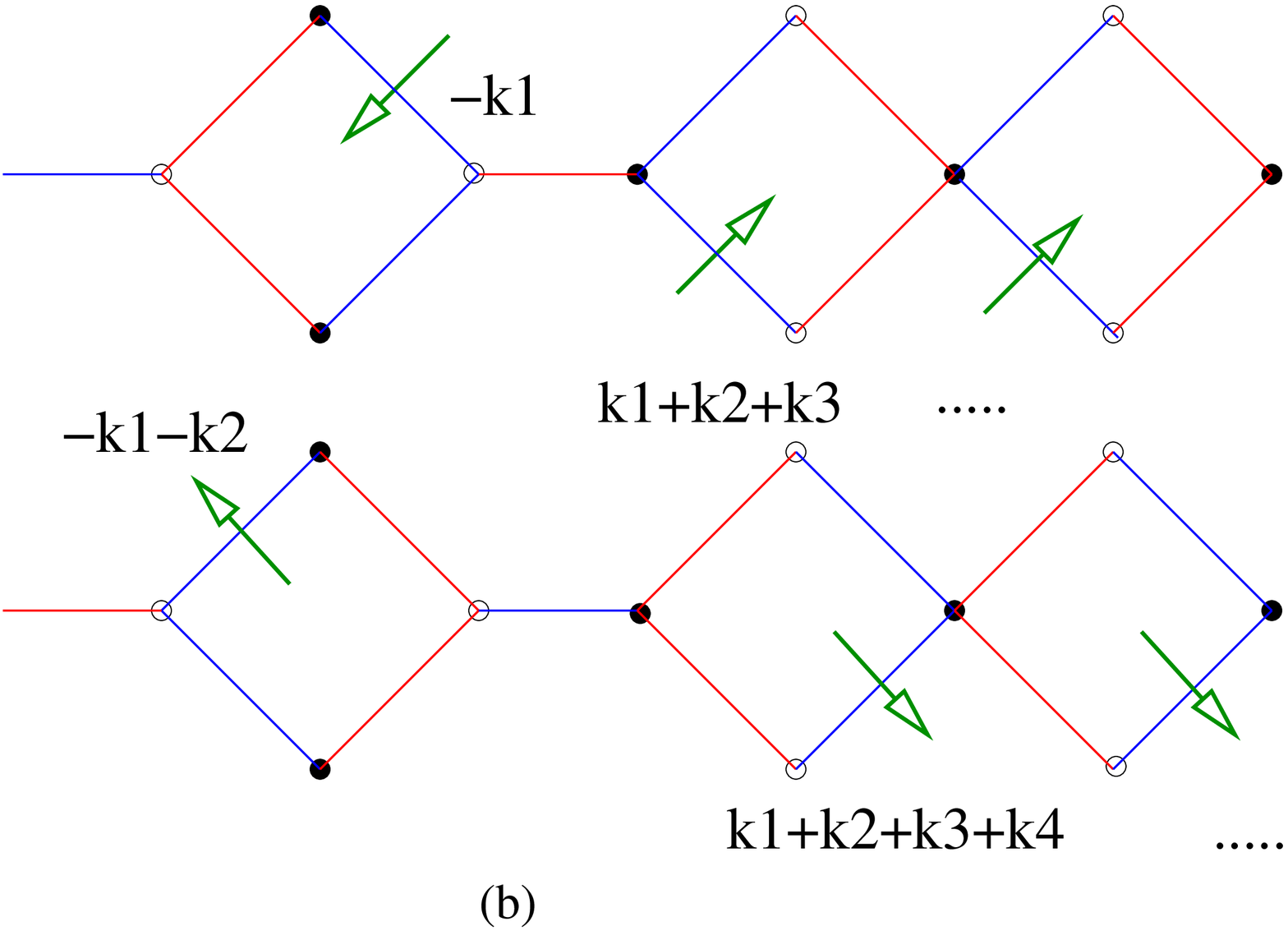}\\
\end{tabular}
\end{center}
\caption{Decomposition of the path $p_3$ on the different perfect matchings}
\label{PM_and_p3}
\end{figure}


\section{Chiral theories}
\label{chiralgeneral}

In this section we study Seiberg like duality for
$\mathcal{N}=2$ three dimensional Chern-Simons theories with 
four dimensional parent chiral theories,
which as such suffer from anomalies.
The anomaly free condition imposes constraints 
on the rank distribution.
In 3d there are no local gauge anomalies.
Nevertheless we work with all $U(N)$ gauge groups
such that the moduli space is the $N$ symmetric
product of the abelian one.
Moreover for three dimensional Chern-Simons 
chiral theories we do not have
 a brane construction as simple as
 for
 the non-chiral case, 
and were not able to deduce the duality from the brane picture.

However, using what we learnt from the non-chiral case, 
 we infer that at least a subset of the possible
three dimensional Seiberg-like toric dualities acts on the field
content and on the superpotential as it does in 4d and moreover
recombines the Chern-Simons levels in a similar way as in
(\ref{dualita}). As a matter of fact, a straightforward extension of
the rule (\ref{dualita}) does not seem to work in the chiral case.
This could be related to the fact that the Master Spaces of the four
dimensional dual Seiberg parents are not isomorphic
\cite{Forcella:2008ng}. For this reason we restrict ourself to the
case where the CS level of the group which undergoes duality is set to
zero.  We assume that the other CS levels are unchanged, and no
fractional branes are created.  This could be suggested by the parity
anomaly matching argument (see appendix \ref{paritya}).  We also set
to zero the CS level of those gauge groups that after duality have the
same interactions with the rest of the quiver as the dualized gauge
group.

By direct inspection we find that under these assumptions on the CS
levels also for chiral 3D CS theories Seiberg like duality leads to
toric duality.  For $\Tilde{\mathbb{F}_0}$ we can take milder
assumptions.  Indeed we find that a generalization of the rule
(\ref{dualita}) to the chiral case still gives toric duality for this
theory.

\subsection{$\Tilde{\mathbb{F}_0}_{\{ k_i \}}$}

Here we study the (2+1)d CS chiral theory whose (3+1)d parent is
$\mathbb{F}_0$.  In (3+1)d there are two dual toric phases of
$\mathbb{F}_0$, denoted as $\mathbb{F}_0^{I}$ and $\mathbb{F}_0^{II}$.
In (2+1)d the two phases for arbitrary choices of the CS levels do not
have the same moduli space.  Nevertheless it is possible to find
assignments for the Chern-Simons levels such that the two phases have
the same toric diagram.

\begin{figure}[h!!!]
\begin{center}
\includegraphics[scale=0.50]{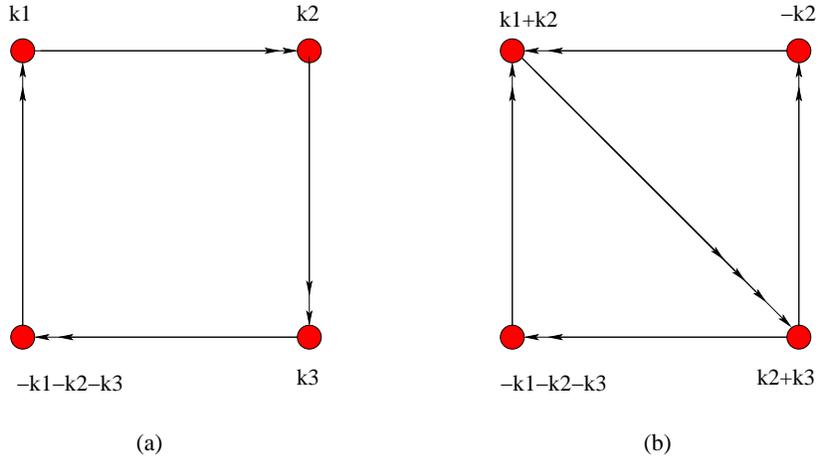}
\caption{(a) quiver for $\Tilde{\mathbb{F}_0^{I}}$ and
(b) quiver for $\Tilde{\mathbb{F}_0^{II}}$ for one of the possible choice of CS levels.}
\label{quivF0}
\end{center}
\end{figure}

The quiver representing the $\Tilde{\mathbb{F}_0^{I}}$ phase 
is in figure \ref{quivF0}(a). The superpotential is 
\be
W = \varepsilon_{ij}\varepsilon_{kl}X_{12}^{i}X_{23}^{k}X_{34}^{j}X_{41}^{l}
\ee
The incidence matrix and the matrix of perfect matchings
are
\be
d=\left(\begin{tabular}{cccccccc}
1&-1&0&0&0&0&-1&1\\
-1&1&1&-1&0&0&0&0\\
0&0&-1&1&1&-1&0&0\\
0&0&0&0&-1&1&1&-1
\end{tabular}\right)
\quad
P=\left(\begin{tabular}{cccccccc}
1 & 1  & 0 & 0  & 0 & 0 & 0 & 0  \\
1 & 0  & 1 & 0  & 0 & 0 & 0 & 0  \\
0 & 0  & 0 & 0  & 1 & 1 & 0 & 0  \\
0 & 0  & 0 & 0  & 1 & 0 & 1 & 0  \\
0 & 1  & 0 & 1  & 0 & 0 & 0 & 0  \\
0 & 0  & 1 & 1  & 0 & 0 & 0 & 0  \\
0 & 0  & 0 & 0  & 0 & 1 & 0 & 1  \\
0 & 0  & 0 & 0  & 0 & 0 & 1 & 1  
\end{tabular}\right)
\ee
The charges of the GLSM fields determine the 
matrix Q, the can be chosen as
\be
Q=\left(\begin{tabular}{cccccccc}
 1&0&0& 0& 0&0&0&-1\\
-1&0&0& 0& 1&0&0& 0\\
 0&0&0& 1&-1&0&0& 0\\
 0&0&0&-1& 0&0&0& 1
\end{tabular}\right)
\ee
The second phase $\Tilde{\mathbb{F}_0^{II}}$ is obtained 
by dualizing node $2$.
The dual superpotential is 
\be
W = \varepsilon_{ij}\varepsilon_{kl} X_{13}^{ik}X_{32}^{l}X_{21}^{j}
  - \varepsilon_{ij}\varepsilon_{kl} X_{13}^{ik}X_{34}^{l}X_{41}^{j}
\ee
The matrices $d$, $P$ and $Q$ are determined from the quiver and the
superpotential
{\tiny{
\be
d=\left(\begin{tabular}{cccc cccc cccc}
-1&-1&0&0&0&0&-1&-1&1&1&1&1 \\
1&1&-1&-1&0&0&0&0&0&0&0&0 \\
0&0&1&1&1&1&0&0&-1&-1&-1&-1 \\
0&0&0&0&-1&-1&1&1&0&0&0&0 
\end{tabular}\right)
\quad
P=\left(\begin{tabular}{cccc cccc c}
0&0&0&0&0&1&0&1&1\\
0&0&1&0&0&1&0&0&1 \\
0&0&0&1&1&0&1&0&0 \\
0&1&0&0&1&0&1&0&0\\
0&0&1&0&0&0&1&0&1\\
0&0&0&0&0&0&1&1&1\\
0&1&0&0&1&1&0&0&0\\
0&0&0&1&1&1&0&0&0\\
1&1&1&0&0&0&0&0&0\\
1&1&0&0&0&0&0&1&0\\
1&0&0&1&0&0&0&1&0\\
1&0&1&1&0&0&0&0&0
\end{tabular}\right)
\ee
}}
\be
Q=\left(\begin{tabular}{cccc cccc c}
1&0&0&0&0&-1&0&0&0\\
0&0&0&0&-1&1&0&0&0\\
-1&0&0&0&0&0&1&0&0\\
0&0&0&0&1&0&-1&0&0
\end{tabular}\right)
\ee

\subsubsection*{Two families}
\label{F0figo}

The $\Tilde{\mathbb{F}_0}_{\{ k_i \}}$ theories turn out to be a special case.
Indeed one can single out two different possibilities: in the first one we put to zero 
just the CS level associated with the group 2; while in the 
second case we can fix to zero just the Chern-Simons level of the group 4 
and transform the CS levels as in the non-chiral case. 
In the first case we choose the CS levels as 
$(k_1,k_2,k_3,k_4)=(k,0,p,-k-p)$.
The CS level matrix for both phases is:
\be
\label{KF0}
C=
\left(
\begin{tabular}{cccc}
1&1&1&1\\
$k$ & $0$ & $p$ & $-k-p$
\end{tabular}
\right)
\ee
and the toric diagram is given by:
\be 
G_t^{(I)}=\left(
\begin{tabular}{cccccccc}
1&1&1&1&1&1&1&1\\
p&0&p&0&p&p&k+p&k+p\\
0&1&-1&0&0&0&0&0\\
0&0&0&0&0&1&-1&0\\
\end{tabular}
\right)
\ee
The CS levels for the dual phase,
obtained by duality on $N_2$,
are unchanged and 
the toric diagram 
\be
G_t^{(II)}=
\left(
\begin{tabular}{ccccccccc}
1&1&1&1&1&1&1&1&1\\
p& k+p & p & p & k+p& k+p& 0 & 0 &0\\ 
0&  0  & -1& 0 & 0  & 0  & 0 & 1 &0\\
0& -1  & 0 & 1 & 0  & 0  & 0 & 0 &0\\
\end{tabular}
\right)
\ee
is equivalent to the one above.

In the second case we choose the CS levels as 
$(k_1,k_2,k_3,k_4)=(k,p,-k-p,0)$. The phase $\Tilde{\mathbb{F}_0^{II}}$ is computed by dualizing 
the node $2$. We observe that by applying the rules 
(\ref{dualita}) the CS levels of $\Tilde{\mathbb{F}_0^{II}}$
are $(k+p,-p,-k,0)$. The $C$ matrices for the two phases are:

\be
C_I=\left(
\begin{tabular}{cccc}
1&1&1&1\\
$k$&$p$&$-k-p$&$0$
\end{tabular}
\right)
\ \ \ \ 
C_{II}=\left(
\begin{tabular}{cccc}
1&1&1&1\\
$k+p$&$-p$&$-k$&$0$
\end{tabular}
\right)
\ee
The toric diagram for the first phase is:
\be 
G_t^{(I)}=\left(
\begin{tabular}{cccccccc}
1&1&1&1&1&1&1&1\\
k& k& 0& 0& k+p& 0& k+p& 0\\ 
0&0&0&0&0&1&-1&0\\
0&-1&1&0&0&0&0&0\\
\end{tabular}
\right)
\ee
while the toric diagram for the second phase is:
\be
G_t^{(II)}=
\left(
\begin{tabular}{ccccccccc}
1&1&1&1&1&1&1&1&1\\
0& 0& 0& k& k& k+p& k& k+p& k+p\\
0&0 & 1 & 0 &0  &0  & 0 & -1 &0\\
0& 1 & 0 & -1 & 0 & 0 & 0 & 0 &0\\
\end{tabular}
\right)
\ee
And they are equivalent.

The $\mathbb{F}_0$ theory seems to be the only 
case where the assumptions
we gave at the beginning of this section
can be relaxed. 
In the following
examples we will just apply those basic rules.

\subsection{$\Tilde{dP_1}_{\{k_i \}}$}
Here we study the (2+1)d CS chiral theory whose (3+1)d parent is
$dP_1$.  In (3+1)d $dP_1$ has only one phase.  After Seiberg duality
on node $2$ the theory has a self similar structure and it is
described by the same quiver.  The only difference is that we have to
change the labels of the groups as ($1 \to 2,2 \to 4,3 \to 1,4 \to
3$).  The matrix $d$, $P$ and $Q$ are unchanged by duality.  The CS
levels in the $C$ matrix change as the labels of the gauge groups do.
We give in figure \ref{antolafai} the two phases.
\begin{figure}[h!!!]
\begin{center}
\includegraphics[scale=0.55]{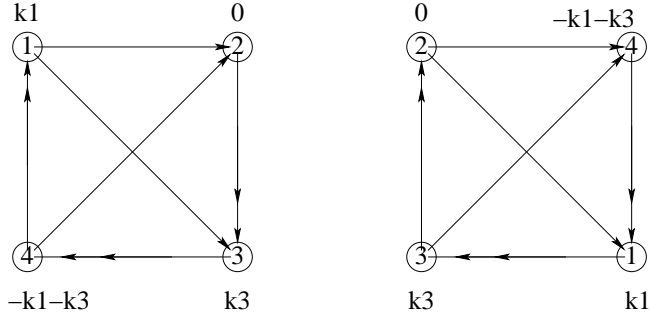}
\caption{
Quiver and CS level for the 
$\Tilde{dP_1}_{\{k_i \}}$ in the two phases
related by duality on node $2$.
}
\label{antolafai}
\end{center}
\end{figure}

We take the assumption described in the introduction of section
\ref{chiralgeneral}.  Hence we choose the CS level of the group that
undergoes duality to be $0$.  With this choice the two phases of the
$2+1$ dimensional theory (see figure \ref{antolafai}) have the same
toric diagram.

The superpotential is 
\be
W= \varepsilon_{ab}X_{13}X_{34}^{a}X_{41}^{b}
   +\varepsilon_{ab}X_{42}X_{23}^{a}X_{34}^{b}
   +\varepsilon_{ab}X_{34}X_{41}^{a}X_{12}X_{23}^{b}
\ee
The $d,P,Q$ matrices are
{\tiny{
\be 
d=\left(
\begin{tabular}{cccccccccc}
1&0&-1&0&0&0&-1&0&0&1\\
0&0&0&0&-1&1&0&1&0&-1\\
-1&1&0&1&0&-1&0&-1&1&0\\
0&-1&1&-1&1&0&1&0&-1&0
\end{tabular}\right)
\quad
 P=\left(
\begin{tabular}{cccccccc}
1&1&1&0&0&0&0&0\\
0&0&0&1&1&1&0&0\\
0&0&0&0&0&0&1&1\\
0&0&0&1&1&0&1&0\\
1&1&0&0&0&0&0&1\\
0&0&1&0&0&1&0&0\\
0&0&0&0&0&1&0&1\\
0&0&1&0&0&0&1&0\\
1&0&0&1&0&0&0&0\\
0&1&0&0&1&0&0&0
\end{tabular}\right)
\ee
}}
\be Q=\left(
\begin{tabular}{cccccccc}
0&0&1&0&1&-1&-1&0\\
0&-1&1&0&0&0&0&0\\
0&0&-1&1&0&0&0&0\\
0&1&-1&-1&-1&1&1&0\\
\end{tabular}\right)
\ee
Following the relabeling of the gauge groups, 
the $C$ matrix in the two phases are
\be
C_1=\left(
\begin{tabular}{cccc}
1&1&1&1\\
$k_1$&0&$k_3$&$-k_1-k_3$
\end{tabular}
\right)
\ \ \ \ 
C_2=\left(
\begin{tabular}{cccc}
1&1&1&1\\
0&$-k_1-k_3$&$k_1$&$k_3$
\end{tabular}
\right)
\ee
The toric diagram for the first theory is
given by the matrix $G_t^{(1)}$
\be
G_t^{(1)}=\left(
\begin{tabular}{cccccccc}
1&1&1&1&1&1&1&1\\
0&0&0&0&0&-1&1&0\\
-1&0&0&0&1&1&0&0\\
$k_1+k_3$&$k_1$&$k_1$&$k_1+k_3$&$k_1$&$k_1$&0&0
\end{tabular}
\right)
\ee
The  toric diagram of the dual theory,
up to $SL(4,Z)$ transformation, is
\be
G_t^{(2)}=\left(
\begin{tabular}{cccccccc}
1 &1&1&1&1& 1&1&1\\
0 &0&0&0&0&-1&1&0\\
-1&0&0&0&1& 1&0&0\\
$k_1+k_3$&$k_1+k_3$&0&$k_1$&$k_1$&0&$k_1$ & $k_1+k_3$
\end{tabular}
\right)
\ee
This shows that the two systems of vectors give
the same toric diagram and that the two
theories have the same abelian moduli
space also in the $2+1$ dimensions,
provided
$k_2=0$.  

\subsection{$\Tilde{dP_2}_{\{ k_i \}}$}
\begin{figure}[h!!!]
\begin{center}
  \includegraphics[scale=0.55]{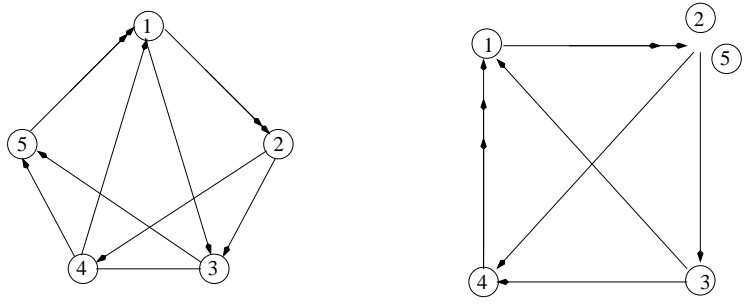}
\caption{The quivers representing the dual phases of $dP2$}
\label{quiverdP2}
\end{center}
\end{figure}

We analyze the (2+1)d CS chiral theory with $dP_2$ as (3+1)d parents.
The $4D$ theory has two inequivalent phases. 
The two phases are connected
by duality on node $5$ and are reported in figure \ref{quiverdP2}.

The constraint on the CS levels explained 
in the introduction of this section 
imposes $k_2=0$ and $k_5=0$. 
Under this assumption the two phases have 
the same toric diagram also for the (2+1)d CS theory.

The superpotential for the two phases are
\bea
W_I &=& X_{13} X_{34} X_{41} - Y_{12} X_{24} X_{41} + X_{12} X_{24} X_{45} Y_{51} - 
  X_{13} X_{35} Y_{51} 
\nonumber\\
&+& Y_{12} X_{23} X_{35} X_{51} - X_{12} X_{23} X_{34} X_{45} X_{51}
\nonumber\\
\nonumber\\
W_{II} &=& Y_{41} X_{15} X_{54} - X_{31} X_{15} X_{53} + Y_{12} X_{23} X_{31} 
- Y_{12} X_{24} X_{41} + Y_{15} X_{53} X_{34} X_{41} \nonumber\\
&-& 
Z_{41} Y_{15} X_{54} + X_{12} X_{24} Z_{41} - X_{12} X_{23} X_{34} Y_{41}
\eea
We compute the toric diagrams
$G_t^{(I)}$ and
$G_t^{(II)}$
for the two phases
\bea
G_t^{(I)}&=&\left(
\begin{tabular}{cccccccccc}
 1&  1& 1& 1& 1& 1& 1&  1& 1 & 1\\
 0&  0& 1& 0& 0& 0& 0& -1& 0& -1\\
 0& -1& 0& 0& 1& 0& 0&  0& 0& -1\\
$k_1$& $k_1$& $k_1$& $k_1$& 0& 0& 0 &0& $-k_3$& $-k_3$
\end{tabular}
\right)
\nonumber \\
G_t^{(II)}&=&\left(
\begin{tabular}{ccccccccccc}
 1&  1& 1& 1& 1& 1& 1&  1& 1 & 1&1\\
 0&  -1& 0& 0& 1& 0& 0& 0& 0& 0&-1\\
 0&  -1& -1& 0& 0& 0& 0& 1& 0& 0&0\\
0&$-k_3$& $k_1$& 0& $k_1$& $-k_3$& $k_1$& 0 &0& 0&0
\end{tabular}
\right)
\eea
They result the same.

\subsection{$\Tilde{dP_3}_{\{ k_i \}}$}

Here we study $\Tilde{dP_3}_{\{ k_i \}}$. This theory has four
phases in four dimensions, with superpotentials
\begin{eqnarray}
W_I &=& X_{13} X_{34} X_{46} X_{61} - X_{24} X_{46} X_{62} 
+ X_{12} X_{24} X_{45} X_{51} -  X_{13} X_{35} X_{51} \nonumber\\
&+ &  X_{23} X_{35} X_{56} X_{62} - X_{12} X_{23} X_{34} X_{45} X_{56} X_{61}  
\non
\end{eqnarray}
\begin{eqnarray}
W_{II} &=& X_{13} X_{34} X_{41} - X_{13} X_{35} X_{51} + X_{23} X_{35} X_{52} 
- X_{26} X_{65} X_{52} + X_{16} X_{65} Y_{51} \nonumber\\ 
&- &  X_{16} X_{64} X_{41} + X_{12} X_{26} X_{64} X_{45} X_{51} 
- X_{12} X_{23} X_{34} X_{45} Y_{51} \non
\end{eqnarray}
\begin{eqnarray}
W_{III} &=& X_{23} X_{35} X_{52} - X_{26} X_{65} X_{52} + X_{14} X_{46} X_{65} Y_{51}
 - X_{12} X_{23} Y_{35} Y_{51} + X_{43} Y_{35} X_{54} \nonumber\\ 
&- &  Y_{65} X_{54} X_{46} + X_{12} X_{26} Y_{65} X_{51} - 
X_{14} X_{43} X_{35} X_{51} \non
\end{eqnarray}
\begin{eqnarray}
W_{IV} &=& X_{23} X_{35} X_{52} - X_{52} X_{26} X_{65} + X_{65} Z_{54} X_{46} 
- Z_{54} X_{41} Y_{15} + Y_{15} Z_{52} X_{21} - Z_{52} X_{23} Y_{35} \nonumber\\
&+ &  Y_{35} X_{54} X_{43} - X_{54} X_{46} Y_{65} + Y_{65} Y_{52} X_{26} 
- Y_{52} X_{21} X_{15} + X_{15} Y_{54} X_{41} - Y_{54} X_{43} X_{35} \non
\end{eqnarray}
Phases (II, III, IV) are computed from phases (I, II, III) 
by dualizing nodes (6, 4, 1) respectively.
The quivers associated with each phase are given in 
Figure \ref{dP3}.
\begin{figure}[h!!!]
\begin{center}
  \includegraphics[scale=0.55]{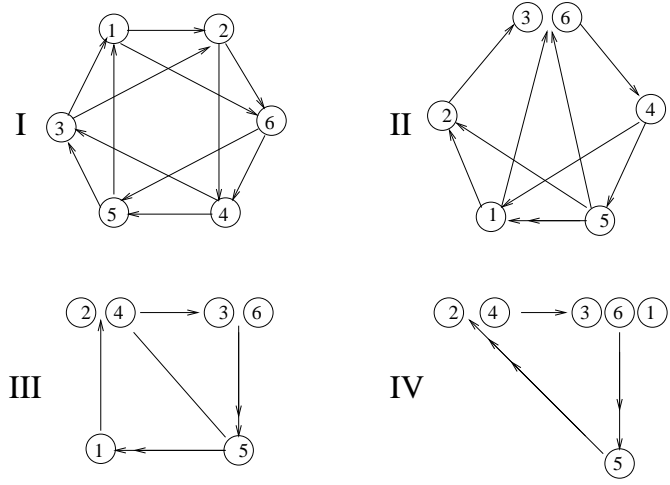}
\caption{The quiver of $dP_3$}
\label{dP3}
\end{center}
\end{figure}
We now show the equivalence of phases (I,II), (II,III) and (III,IV) 
by choosing $(k_3=k_6=0)$, $(k_2=k_4=0)$ and $(k_1=k_3=k_6=0)$ 
respectively.
For phases $(I,II)$ we have
\tiny
\be
G_t^{(I)}=\left(
\begin{tabular}{cccccccccccc}
 1 &  1 &  1 &  1 &  1 &  1 &  1 &  1 &  1 &  1 &  1 &  1 \\
-1 &  0 &  0 &  0 &  0 &  1 &  0 &  0 &  0 & -1 &  1 &  0 \\
-1&   0 & -1 &  0 &  0 &  1 &  0 &  1 &  0 &  0 &  0 &  0 \\
$ k_1 + k_2$&0&$k_1$&$k_1+k_2+k_4$&$k_1+k_2$&0&$k_1$&$k_1+k_2+k_4$&$k_1+k_2$& 
$k_1 + k_2$&  0 &  0 \\
\end{tabular}
\right)
\nonumber
\ee
\be
G_t^{(II)}=\left(
\begin{tabular}{ccccccccccccc}
 1 &  1 &  1 &  1 &  1 &  1 & 1 &  1 & 1 & 1 &  1 &  1 & 1 \\
 0 &  0 & -1 &  0 &  0 &  0 & 0 & -1 & 0 & 0 &  0 &  1 & 1 \\
 0 &  1 &  0 &  0 &  0 & -1 & 0 & -1 & 0 & 0 &  0 &  0 & 1 \\
 $k_1$  & $k_1 + k_2 + k_4$ & $k_1 + k_2$ & $k_1 + k_2$ & $k_1 + k_2$ & 
 $k_1$  & $k_1 + k_2 + k_4$ & $k_1 + k_2$ & $k_1 + k_2$ & $k_1 + k_2$ &  
 0&   0 &  0  
\end{tabular}\nonumber
\right)
\ee
\normalsize 
For phases $(II,III)$ we have
\tiny
\be
G_t^{(II)}=\left(
\begin{tabular}{ccccccccccccc}
 1 &  1 &  1 &  1 &  1 &  1 &  1 &  1 &  1 &  1 &  1 &  1 &  1 \\
 0 &  1 &  1 &  0 &  0 & -1&  0 &  0 &  0 &  0 &  0 & -1&0 \\
 0 &  0 &  1 &  0 &  0 &  0 &  0 &  1 &  0 &  0 &    0 & -1& -1\\
$-k_1$& $k_3 + k_5$& $k_3 + k_5$& $k_3 + k_5$& $-k_1$&$-k_1 - k_3$&$k_5$&
$-k_1 - k_3$&$k_5$&$-k_1 - k_3$& 
0& 0& 0
\end{tabular}\right)\nonumber
\ee
\be
G_t^{(III)}=\left(
\begin{tabular}{cccccccccccccc}
 1 &  1 &  1 &  1 &  1 &  1 &  1 &  1 &  1 &  1 &  1 &  1 &  1 & 1 \\
 0 &  1 &  0 &  0 &  1 &  0 & -1 &  0 &  0 & -1 &  0 &  0 &  0 & 0 \\
 0 &  1 & -1 &  0 &  0 &  0 & -1&  0 &  0 &  0 &  0 &  0 &  1 & 0 \\
 0 & $k_3+k_5$ &  0 & $-k_1$& $k_3+k_5$ &$-k_1$ &  0 & $k_3+k_5$ & 
 $k_5$ & $-k_1-k_3$& $-k_1$&$-k_1$&$-k_1-k_3$& $-k_1-k_3$\\
\end{tabular}\nonumber
\right)
\ee
\normalsize 
For phases $(III,IV)$ we have
\tiny
\be
G_t^{(III)}=\left(
\begin{tabular}{cccccccccccccc}
 1 &  1 &  1 &  1 &  1 &  1 &  1 &  1 &  1 &  1 &  1 &  1 &  1 &  1 \\
 0 & -1&  1 &  0 &  0 &  0 &  1 &  0 &  0 &  0 &  0 &  0 & -1 &  0 \\
  0 &-1&  0 &  0 & -1&  0 &  1 &  0 &  0 &  1 &  0 &  0 &  0 &  0 \\
$k_2+k_4$& $k_2$& $k_2+k_4$& $k_2$& $k_2+k_4$& $k_2+k_4$& 
 $k_4$ &  0 &  0 &  0 &  0 &  $k_4$ &  0 &  0
\end{tabular}\nonumber
\right)
\ee
\be
G_t^{(IV)}=\left(
\begin{tabular}{ccccccccccccccccc}
1& 1& 1& 1& 1& 1& 1& 1& 1& 1& 1& 1& 1&1& 1& 1&1\\
0& 0& 0& 0& 1& 0& 0& 0& 0& 0& 0& 1& 0& -1& 0& 0& -1\\
0& 0& 0& 0& 0& -1& 0& 0& 0& 0& 0& 1&  0& -1& 1& 0& 0\\
$k_2 + k_4$&
$\!\!\!\!\!\!k_2 + k_4$&
$\!\!\!\!\! k_2 + k_4$& $\!\!\!\!\!k_2 + k_4$& 
$\!\!\!\!\! k_2 + k_4$& $\!\!\!\!\!k_2 + k_4$& 
$\!\!\!\!\! k_2 + k_4$& $\!\!\!\!\!k_2 + k_4$& $\!\!\!\!\!k_2 + k_4$& 
$k_2 + k_4$& $k_4$& 
$k_4$& 
$k_2$& 
$k_2$& 0& 0& 0
\end{tabular}\nonumber
\right)
\ee

\normalsize 

\subsection{$\Tilde{Y^{32}}_{\{ k_i \}}$}
This is the last chiral theory we analyze. In this case, after duality,
there is not an identification between the gauge group that undergoes
duality with other groups.  This implies that the assumptions of
section \ref{chiralgeneral} impose only $k_g=0$, where $k_g$ is the CS
level of the dualized gauge group.  As for the case of $dP_1$ and all
the $Y^{p,p-1}$ theories, the $Y^{3,2}$ theory is self similar under
four dimensional Seiberg duality.  We can evaluate $\mathcal{M}_4$ for
one phase and then the toric diagram associated with a dual phase is
given by an appropriate change of the $D$-term modding matrix.

We fix the conventions on the groups by giving the tiling of
the two dual phases, see Figure \ref{tilingY32}.
The two phases are connected by duality on node $5$,
so we set $k_5=0$.

\begin{figure}[h!!!]
\begin{center}
  \includegraphics[scale=0.55]{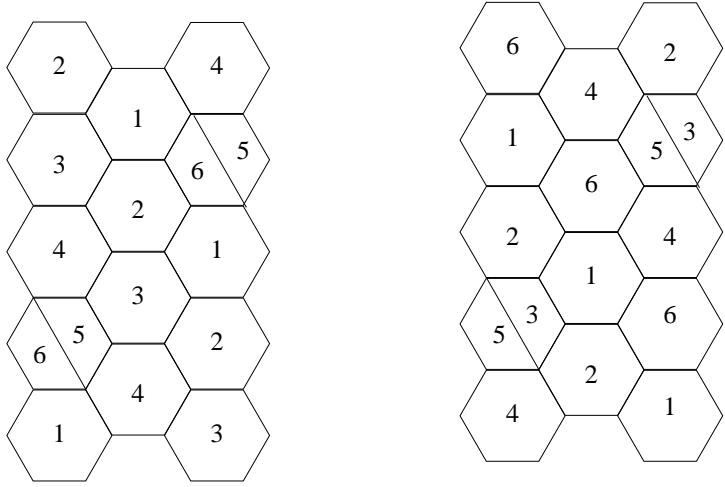}
\caption{The tiling for the two dual phases of $Y^{32}$. 
Seiberg duality has been performed on groups $5$.}
\label{tilingY32}
\end{center}
\end{figure}
The CS matrices for the two dual phases are: 
\footnotesize
\be
C_1=\left(
\begin{tabular}{cccccc}
1&1&1&1&1&1\\
$k_{1}$&$k_{2}$&$k_{3}$&$k_{4}$&0&$-k_1-k_2-k_3-k_4$
\end{tabular}
\right)
\ \ \ \ \ 
C_2=\left(
\begin{tabular}{cccccc}
1&1&1&1&1&1\\
$k_{4}$&$-k_1-k_2-k_3-k_4$&$k_1$&$k_2$&$k_3$&0
\end{tabular}
\right)\nonumber
\ee
\normalsize
The toric diagrams are encoded in the $G_t$ matrices.
\tiny
\be
G_t^{(1)}=
\left(
\begin{tabular}{cccccccccccccccccc}
1&1&1&1&1&1&1&1&1&1&1&1&1&1&1&1&1&1\\
0&0&0&0&0&0&0&0&0&0&0&0&1&0&0&0&0&-1\\
2&1&1&1&0&1&0&0&1&0&0&0&-1&-1&1&0&0&0\\
$k_3+k_4\!\!\!\!\!\!$&0&$\!\!\!\!\!\!-k_6$&$\!\!\!\!\!\!k_2+k_3+k_4$&
$\!\!\!\!\!\!k_2$&$\!\!\!\!\!\!k_3+k_4$&0&$-k_6$&$k_4$&$-k_3$&
$\!\!\!\!\!\!k_1+k_2+k_4$&$\!\!\!\!\!\!k_2+k_4$&$\!\!\!\!\!\!k_1+2k_2+k_4$&$\!\!\!\!\!\!k_2-k_3$&0&$k_2$&0&0
\end{tabular}
\right)\nonumber
\ee
\be
G_t^{(2)}=\left(
\begin{tabular}{cccccccccccccccccc}
1&$\!\!\!\!\!\!$1&$\!\!\!\!\!\!$1&$\!\!\!\!\!\!$1&$\!\!\!\!\!\!$1&$\!\!\!\!\!\!$1&$\!\!\!\!\!\!$1&1&1&1&1&1&1&1&1&$\!\!\!\!\!\!$1&$\!\!\!\!\!\!$1&$\!\!\!\!\!\!$1\\
0&$\!\!\!\!\!\!$0&$\!\!\!\!\!\!$0&$\!\!\!\!\!\!$0&$\!\!\!\!\!\!$0&$\!\!\!\!\!\!$0&$\!\!\!\!\!\!$0&0&0&0&0&0&1&0&0&$\!\!\!\!\!\!$0&$\!\!\!\!\!\!$0&$\!\!\!\!\!\!$-1\\
2&$\!\!\!\!\!\!$1&$\!\!\!\!\!\!$1&$\!\!\!\!\!\!$1&$\!\!\!\!\!\!$0&$\!\!\!\!\!\!$1&$\!\!\!\!\!\!$0&0&1&0&0&0&-1&-1&1&$\!\!\!\!\!\!$0&$\!\!\!\!\!\!$0&$\!\!\!\!\!\!$0\\
$\!\!\!\!\!\! k_3+k_4$&
$\!\!\!\!\!\!k_4$&
$\!\!\!\!\!\!k_4$&
$\!\!\!\!\!\!$0&
$\!\!\!\!\!\!-k_3$&
$\!\!\!\!\!\!-k_6$&
$\!\!\!\!\!\!k_1+k_2+k_4$&
$\!\!\!\!\!\!k_1+k_2+k_3$&
$\!\!\!\!\!\!k_2+k_3+k_4$&
$\!\!\!\!\!\!k_2+k_4$&
$\!\!\!\!\!\!k_2+k_4$&
$k_2$&
$\!\!\!\!\!\!k_1+2k_2+k_4$&
$\!\!\!\!\!\!k_2-k_3$&$k_3+k_4$&
$\!\!\!\!\!\!$0&
$\!\!\!\!\!\!$$-k_6$&
$\!\!\!\!\!\!$0
\end{tabular}
\right)\nonumber
\ee
\normalsize
$\!\!\!\!\!\!$
and they coincide for arbitrary $k_1,k_2,k_3,k_4$, remind $k_6=-k_1-k_2-k_3-k_4$.

\section{Dualities for CS theories without 4d parents}

Three dimensional CS theories with four dimensional parents are a
subset of all the possible 3d CS theories \cite{Franco:2008um}.  For
CS theories without four dimensional parents we miss in principle the
intuition from the 4d Seiberg duality. In this short section we see
that we can still describe a subset of 3d CS theories with the same
mesonic moduli space if we just apply the rules we learnt in the previous
sections.

We study a case associated with $Q^{111}$. We show that by performing
a Seiberg-like duality and by setting the CS level of the dualized
gauge group to zero, the toric diagrams of the two models coincide.

\begin{figure}[h!!!]
\begin{center}
  \includegraphics[scale=0.55]{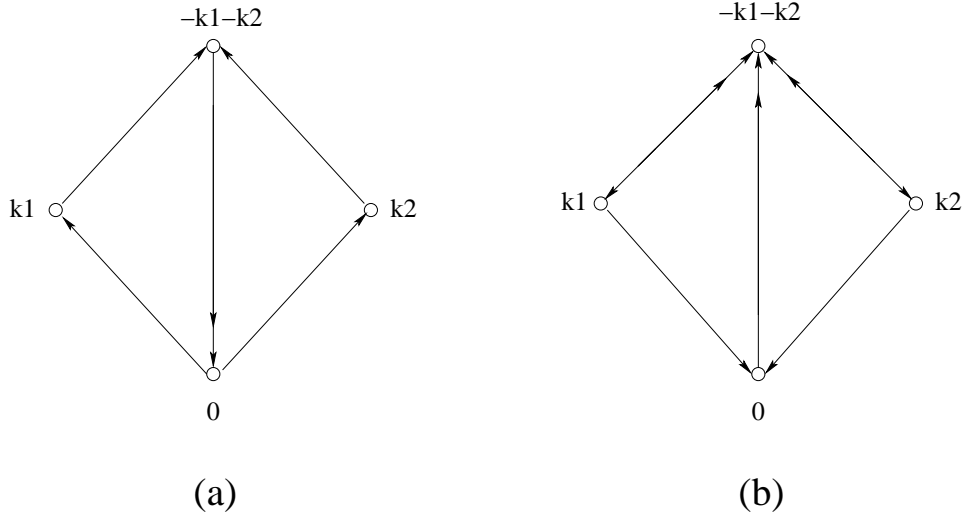}
\caption{The quiver for $Q^{111}$ in the two dual phases.}
\label{Q111quiv}
\end{center}
\end{figure}

\subsection{Example}
The theory is described by the quiver given in Figure \ref{Q111quiv}a.
It is a generalization of the $C(Q_{1,1,1})$ with arbitrary CS levels \cite{Franco:2008um}.
The superpotential is 
\be
W= X_{41}X_{13}X_{34}^{1}X_{42}X_{23}X_{34}^{2}
-
X_{41}X_{13}X_{34}^{2}X_{42}X_{23}X_{34}^{1}
\ee
The toric diagram for $k_4=0$ is given by
\be
G_t = \left(\begin{tabular}{cccccc}
1&1&1&1&1&1\\
1&0&1&0&1&0\\
1&0&0&1&1&0\\
$k_1$&$k_2$&0&0&0&0
\end{tabular}
\right)
\ee
Seiberg duality on node $N_4$ gives the superpotential
\bea
W&=& 
X_{13}X_{32}^{(1)}X_{23}X_{31}^{(2)} 
- 
X_{13}X_{32}^{(2)}X_{23}X_{31}^{(1)}
+
X_{24}X_{43}^{(2)}X_{32}^{(2)}
-
X_{24}X_{43}^{(1)}X_{32}^{(1)}
\nonumber\\
&+&
X_{14}X_{43}^{(1)}X_{31}^{(1)}
-
X_{14}X_{43}^{(2)}X_{31}^{(2)}
\eea
and the theory is described by the quiver given in Figure \ref{Q111quiv}b.
The toric diagram in this case is
\be
G_t = \left(\begin{tabular}{cccccccc}
1&1&1&1&1&1&1&1\\
1&0&1&0&1&0&1&0\\
1&0&1&0&1&1&0&0\\
$k_1$&$k_2$&$k_1$&$k_2$&0&0&0&0
\end{tabular}
\right)
\ee
and it is equivalent to the toric diagram for the first phase.
\section{Conclusion}

In this paper we have some advances towards the understanding 
of toric duality for M$2$ branes. Generalizing the work of 
\cite{Giveon:2008zn,Aharony:2008gk}, we proposed a Seiberg-like duality
for non-chiral three dimensional CS matter theories and 
we verified that the mesonic moduli space of dual theories is indeed
the same four-fold Calabi Yau probed by the M$2$ branes. 
In the chiral case and in the case in which the three dimensional 
theories do not have a four dimensional parent, the situation 
is more complicated. However, fixing to zero the value of
some of the Chern-Simons levels, we were able to realize toric dual pairs.

We have just analyzed the mesonic moduli space, it would be important
to study the complete moduli space, including baryonic operators.
 
For the non-chiral case the two main limitations are the lack of understanding 
of the transformation rule for the rank of the gauge groups and the fact 
that we forced to zero some of the $k_i$.
It is reasonable that there exist some more general and precise transformation 
rules and we would like to investigate them. 

We concentrated on Seiberg-like transformations, but it is well known that 
in three dimension there exist duality maps that change the number 
of gauge group factors. It would be interesting to systematically study these 
more general transformations.

A lot of possible directions and generalizations are opening up 
and after these first steps we hope to step up.

\section*{Acknowledgments}
We are happy to thank Alberto Zaffaroni for many nice discussions and 
Ami Hanany 
for comments.
We are grateful to the authors of \cite{Franco,Ami}
for informing us about their research on similar topics.

A.~A.~ and L.~G.~ are supported in part by INFN,
in part  by
MIUR under contract 2007-5ATT78-002
and in part by 
the European
Commission RTN programme MRTN-CT-2004-005104.
D. ~F.~ is supported by CNRS and ENS Paris.
A. ~M. ~ is
supported in part 
by the Belgian Federal Science Policy Office 
through the Interuniversity 
Attraction Pole IAP VI/11, by the European
Commission FP6 RTN programme MRTN-CT-2004-005104 and by 
FWO-Vlaanderen through project G.0428.06.

\appendix

\section{Parity anomaly}\label{paritya}
We briefly review parity anomaly for 3D gauge theories \cite{AlvarezGaume:1983ig, Redlich:1983dv}
and the parity anomaly matching agument \cite{Aharony:1997bx}.
In three dimensions there are no local gauge anomalies. However gauge invariance can require the introduction
of a classical Chern-Simons term, which breaks parity. This is referred to as parity anomaly.
 
For abelian theories with multiple $U(1)$'s, 
there is a parity anomaly
if
\be
\mathcal{A}_{ij}=\frac{1}{2} \sum_{\text{fermion}} (q_f)_i (q_f)_j 
\quad \in \quad  \mathbb{Z}+\frac{1}{2}
\ee
Here $(q_f)_i$ is the charge of the fermion $f$ under the $U(1)_i$. 
We work in a basis where all the charges are integers.

\subsection*{Parity anomaly matching}
In the context of dualities in 4D gauge theory, a relevant tool have been the 
t'Hooft anomaly matching between the electric and the magnetic description.
Having some global symmetries, we suppose that they are gauged and we compute
their anomaly. The result of this computation should be equal in the two dual
descriptions.
The same technique can be used here for the parity anomaly. We suppose we gauge
the global $U(1)$'s of the theory, and we compute their parity 
anomaly both in the electric and in the
magnetic description. 
The two computations should match.

The parity anomaly matching is much weaker then the t'Hooft one.
Indeed in 4D the precise anomalies associated with gauging global
symmetries must match. In 3D there is a weaker $\mathbb{Z}_2$ type condition. 

For the $\Tilde{L^{aba}}_{\{k_i\}}$ theories, parity anomaly matching is 
obeyed by the Seiberg like duality we propose.

In chiral theories, parity anomaly matching can be non trivial
between dual phases 
if fractional branes are introduced by the duality.

As an example we analyze the toric 
chiral (2+1)d CS theory which has as (3+1)d parents $dP_2$.
We consider
the parity anomaly associated
to the two $U(1)$ flavour symmetries $F_1$ and $F_2$.
The charges of the chiral fields of the theory 
under these symmetries 
can be derived from
\cite{Forcella:2008ng}.  
The electric theory has equal ranks $n$.
In the magnetic description we set rank $n+k$ for the dualized gauge group (number 5)
 and $n$ for the others. The integer $k$ counts the number of fractional branes
 introduced in the duality.
The parity anomaly matrices are
 \be
  \mathcal{A}_{ele}
  \in
 \left(
 \begin{array}{ccc}
 \mathbb{Z} &  \mathbb{Z} \\
 \mathbb{Z} & \mathbb{Z}
  \end{array}
 \right)
 \ee
and
 \be
 \mathcal{A}_{mag}
 \in
 \left(
 \begin{array}{ccc}
 \mathbb{Z} & \mathbb{Z}+\frac{nk}{2} \\
 \mathbb{Z}+\frac{nk}{2} & \mathbb{Z}
  \end{array}
 \right)
 \ee
One can see that 
in the electric description there are no parity anomalies. In the magnetic description the
 off diagonal components of $\mathcal{A}_{mag}$ can instead lead to parity anomaly
 if $k n$ is odd. 
 If we set $k=0$ the electric and magnetic theories
 satisfies the parity anomaly matching for the two flavour symmetries.

\end{document}